\documentclass[format=acmsmall, review=false, screen=true, anonymous=false]{acmart}

\usepackage[utf8]{inputenc}
\usepackage[english]{babel}

\usepackage{booktabs} 

\usepackage[ruled]{algorithm2e} 

\SetAlFnt{\small}
\SetAlCapFnt{\small}
\SetAlCapNameFnt{\small}
\SetAlCapHSkip{0pt}
\IncMargin{-\parindent}

\usepackage[export]{adjustbox} 
\usepackage{amssymb,amsmath}

\usepackage{enumitem}

\usepackage{calc}

\usepackage{multirow}

\usepackage{siunitx}
\usepackage{makecell}

\definecolor{VeryLightGray}{rgb}{0.92,0.92,0.92}
\definecolor{VeryVeryLightGray}{rgb}{0.97,0.97,0.97}
\usepackage{colortbl}
\newcommand{\graybg}{\cellcolor{VeryLightGray}}
\newcommand{\lightgraybg}{\cellcolor{VeryVeryLightGray}}

\setcopyright{acmlicensed}
\acmJournal{PACMHCI}
\acmYear{2018}
\acmVolume{2}
\acmNumber{CSCW}
\acmArticle{166}
\acmMonth{11}
\acmPrice{15.00}
\acmDOI{10.1145/3274435}

\begin{document}
\title[Constructing Urban Tourism Space Digitally]{Constructing Urban Tourism Space Digitally: A Study of Airbnb Listings in Two Berlin Neighborhoods}

\author{Natalie Stors}
\affiliation{%
  \institution{Tourism Geography, University of Trier}
  \city{Trier}
  \country{Germany}
}
\email{stors@uni-trier.de}

\author{Sebastian Baltes}
\affiliation{%
  \institution{Software Engineering, University of Trier}
  \city{Trier}
  \country{Germany}
}
\email{research@sbaltes.com}


\begin{abstract}
Over the past decade, Airbnb has emerged as the most popular platform for renting out single rooms or whole apartments.
The impact of Airbnb listings on local neighborhoods has been controversially discussed in many cities around the world.
The platform's widespread adoption led to changes in urban life, and in particular urban tourism.
In this paper, we argue that urban tourism space can no longer be understood as a fixed, spatial entity.
Instead, we follow a constructionist approach and argue that urban tourism space is (re-)produced digitally and collaboratively on online platforms such as Airbnb.
We relate our work to a research direction in the CSCW community that is concerned with the role of digital technologies in the production and appropriation of urban space and use the concept of representations as a theoretical lens for our empirical study.
In that study, we qualitatively analyzed how the two Berlin neighborhoods \emph{Kreuzkölln} and \emph{City West} are digitally constructed by Airbnb hosts in their listing descriptions.
Moreover, we quantitatively investigated to what extend mentioned places differ between Airbnb hosts and \emph{visitBerlin}, the city's destination management organization (DMO).
In our qualitative analysis, we found that hosts primarily focus on facilities and places in close proximity to their apartment. 
In the traditional urban tourism hotspot \emph{City West}, hosts referred to many places also mentioned by the DMO.
In the neighborhood of \emph{Kreuzkölln}, in contrast, hosts reframed everyday places such as parks or an immigrant food market as the must sees in the area.
We discuss how Airbnb hosts contribute to the discursive production of urban neighborhoods and thus co-produce them as tourist destinations.
With the emergence of online platforms such as Airbnb, power relations in the construction of tourism space might shift from DMOs towards local residents who are now producing tourism space collaboratively.
\end{abstract}

%
%


%
%


\maketitle


\section{Introduction}

Airbnb, the largest digital peer-to-peer distribution platform for accommodations, operates in about 81,000 cities in 191 countries, offering 4.5 million rooms, apartments, houses, and other types of accommodation~\cite{AirbnbInc2018}.
The widespread adoption of Airbnb led to changes in urban life and in particular urban tourism.
Visitors of urban destinations increasingly leave inner-city areas that are close to major sights and tourism-related facilities, and venture into residential neighborhoods---a phenomenon called \emph{off the beaten track} or \emph{new urban tourism}~\cite{PappaleporeMaitlandOthers2010, PappaleporeMaitlandOthers2014, FullerMichel2014, StorsKagermeier2017, MaitlandNewman2009}.
Short-term rentals seem to foster this development~\cite{FullerMichel2014, FreytagBauder2018, IoannidesRoeslmaierOthers2018}.
They enable all kinds of temporary city-users~\cite{Martinotti1999}, such as cultural tourists, exchange students, temporary migrants, or business travelers, to stay in private apartments.
However, only a fraction of urban residents participates in the practice of renting out rooms or apartments.
Some of them are annoyed by the large influx of visitors to their neighborhood and the consequences for the urban structure resulting from large-scale short term rentals~\cite{Gant2016, ColombNovy2016, NofreGiordanoOthers2017}.
The impact of the fast-growing number of Airbnb listings, particularly in residential neighborhoods, has been controversially discussed in many cities around the world.
Central topics of this debate were, among others, Airbnb’s contribution to the transformation of residential neighborhoods and resulting gentrification processes~\cite{GravariBarbasGuinand2018, Gant2016}, coming along with rent increase~\cite{Mermet2018, SchaferHirsch2017}, changes in the social structure of the neighborhood~\cite{Gant2016, SansQuaglierei2016}, and effects on the hotel industry~\cite{ZervasProserpioOthers2017}.
Legal issues, such as the prohibition of Airbnb in some cities, or restrictions of its offer in others, were likewise central points of discussion~\cite{DredgeGyimothy2017, GuttentagSmithOthers2017, QuattroneProserpioOthers2016}.

In order to fully understand the Airbnb phenomenon and its consequences for the city, it is important to consider both the tourist and the host perspective.
Tourists and their spatial practices have already been considered in academic research from urban geography and the CSCW research community.
While Pappalepore, Maitland, and Smith were primarily concerned with visitors' motivation for leaving central tourist areas~\cite{PappaleporeMaitlandOthers2010, PappaleporeMaitlandOthers2014}, Brown and Perry looked into tourists' usage of maps~\cite{BrownPerry2001}.
Researchers also discussed the potential of electronic tourist guides~\cite{CheverstDaviesOthers2000, KenterisGavalasOthers2009} and in particular mobile tourism recommendation systems~\cite{GavalasKenteris2011}.
Studies on mobility practices provided insights into visitors' spatio-temporal flows throughout the city and highlighted the significance of mobile technologies such as GPS trackers as new tools for engaging in this research field~\cite{BirenboimShoval2016, GrinbergerShovalOthers2014, ShovalAhas2016}.
Digital information technologies proved helpful in identifying tourists' mobility practices even outside central tourist areas~\cite{Bauder2015, FreytagBauder2018}.


Local communities, in contrast, have long been neglected in urban tourism research.
They were mainly described as suffering from urban transformation processes, rent increase, and gentrification~\cite{Novy2011, FullerMichel2014}. 
However, some researchers pointed to the significance of local people in the co-construction of urban tourism space and atmospheres~\cite{PappaleporeMaitlandOthers2014, RussoRichards2016}.
Airbnb hosts are increasingly gaining attention in this context.
Studies exist that look into hosts' motivation to participate in online hospitality networks~\cite{StorsKagermeier2017, LampinenCheshire2016}.
Such research approaches deal with drivers for renting out~\cite{LampinenCheshire2016, StorsKagermeier2017, Ke2017} and analyze the importance of monetary transactions in hospitality exchange networks~\cite{IkkalaLampinen2015}.
Other studies were concerned with hosts' trustworthiness~\cite{ErtFleischerOthers2016, MaHancockOthers2017}.
What has so far rarely been taken into account is hosts’ potential to contribute to the discursive and performative reframing of residential neighborhoods into urban tourism areas, which is the focus of our work.

Generally, user-generated content on information sharing platforms such as TripAdvisor has been identified as an important source of information for tourists~\cite{XiangGretzel2010}.
Regarding the implications that online reviews provide for urban space, Zukin et al. found that Yelp reviews produce positive or negative space images and thus may contribute to economic (dis)investment in urban areas~\cite{ZukinLindemanOthers2015}.
Similarly, Corbett outlined how space images are affected by descriptions on an online real estate information platform~\cite{Corbett2017}.
With regard to the CSCW community, he argues that the concept of place has been widely adapted to inform location-based technologies, whereas implications of a \emph{``discursive place shaping''} on online platforms have not been considered yet.
With our research, we try to fill this gap.
The overarching goal of our research is to investigate:

\begin{quote}
\emph{How are online platforms engaging in the co-production of (new) urban tourism space?}
\end{quote}

In particular, we are interested in how the peer-to-peer accommodation platform Airbnb contributes to shaping tourist places in the city.
In Section~\ref{sec:theoretical-background}, we embed this research direction in varying conceptualizations of space and place as applied in the CSCW and tourism geography research communities, and discuss the common ground between their research perspectives.

We identify that container-like understandings of urban tourism space~\cite{Framke2002}, as represented in the \emph{tourist-historic city model}~\cite{AshworthHaan1985, AshworthTunbridge1990, AshworthTunbridge2000}, resemble the traditional framing of space as a \emph{``natural fact''}~\cite{HarrisonDourish1996} in the CSCW community.
Both approaches, however, are lacking satisfying explanations of how tourism space can emerge and develop in residential neighborhoods.
The \emph{tourist-historic city model} solely takes into account tourism-related facilities and infrastructure as defining elements of tourism space.
Such structures are often non-existing in new urban tourism areas like \emph{Kreuzkölln}, one of our case-study neighborhoods.
Still, this neighborhood without any major sights is becoming a tourism hotspot~\cite{FullerMichel2014, ColombNovy2016}.
Against this background, we follow a constructionist understanding of urban tourism space~\cite{Young1999, Framke2002, Iwashita2003}.
We argue that tourism space, like tourist sights~\cite{MacCannell1976}, is socially constructed through \emph{representations}~\cite{PritchardMorgan2000, Saarinen2004} and \emph{performances}~\cite{Edensor1998, Edensor2000, Edensor2001, BrenholdtHaldrupOthers2004, Larsen2008}.
That means we no longer regard tourism facilities and infrastructure as the central elements defining tourism space.
Instead, people are the major agents who transform places and landscapes into tourist destinations.
They attach meanings and values to places and objects~\cite{Davis2001, Saarinen2004}, produce written, oral, or pictorial \emph{representations} of them, and thus contribute to the discourse of how places or objects are to be perceived.
Finally, following the performative turn in tourism studies~\cite{Larsen2008, Larsen2012}, we argue that places need to be enacted through \emph{``bodily performances''}~\cite{Larsen2012, BrenholdtHaldrupOthers2004}.
Practices, such as picture taking or collectively \emph{``gazing''}~\cite{Urry1990} upon a building, are necessary to enact places in a touristic manner~\cite{Larsen2014}. 
Taking these theoretical considerations into account, we analyze how two different Berlin neighborhoods, \emph{Kreuzkölln} and \emph{City West}, are socially constructed in Airbnb listings.
The following three questions guided our research:

\begin{itemize}[labelindent=\parindent, labelwidth=\widthof{\textbf{RQ1:}}, label=\textbf{RQ1:}, leftmargin=*, align=parleft, parsep=0pt, partopsep=0pt, topsep=1ex, noitemsep]
\item[\textbf{RQ1:}] How are the two neighborhoods \emph{Kreuzkölln} and \emph{City West} constructed as tourism spaces in Airbnb listings?
\item[\textbf{RQ2:}] How does the space construction differ between these two neighborhoods?
\item[\textbf{RQ3:}] How do the neighborhood descriptions differ between Airbnb hosts and the destination management and marketing organization (DMO)?
\end{itemize}

We collected the listing descriptions from all Airbnb listings located in our research areas, resulting in a total number of 960 descriptions (see Section~\ref{sec:data-collection}).
Afterwards, we randomly selected 100 listings that we qualitatively analyzed, applying grounded theory coding techniques~\cite{Charmaz2014, Saldana2015} (see Section~\ref{sec:data-analysis}).
Our goal was to identify which elements of the neighborhoods are mentioned in the listings and how they are described as being touristically significant.
Moreover, we analyzed which practices are encouraged in the listings (see Sections~\ref{sec:places-facilities},~\ref{sec:streets-squares},~\ref{sec:sights-parks-markets}) and investigated how places named by Airbnb hosts and \emph{visitBerlin}, the city's destination management and marketing organization (DMO), differ (see Sections~\ref{sec:quantitative-analysis}~and~\ref{sec:comparison-dmo}).

This paper provides a twofold contribution:
From a theoretical perspective, a container-like notion of physical tourism space is overcome by understanding space as being socially constructed.
This theoretical reframing of space is necessary to be able to explain how residential neighborhoods that are lacking any sights can become tourist places.
What is particularly new is the empirical focus on the digital and collaborative construction of tourism space in Airbnb listings.
Traditionally, the DMO produced promotional space images and steered visitors' attention and practices.
After the proliferation of digital technologies and in particular peer-to-peer platforms, Airbnb hosts are now able to participate in the discursive framing of their neighborhood---they gained the ability to (re)interpret space and endow neighborhoods with a touristic meaning.

\section{Theoretical Background}
\label{sec:theoretical-background}

Understanding \emph{space} and \emph{place} on a conceptual level is a central goal of geography.
Much empirical work, however, deals with natural or cultural phenomena happening in distinct \emph{places}.
Considering the nature of \emph{space} just became popular again after the \emph{``spatial turn''}~\cite{Soja1989} in social sciences from the late 1960s onwards~\cite{Soja2009}.
This development induced debates on the nature of space and place in various disciplines, including the CSCW research community~\cite{HarrisonDourish1996, FitzpatrickKaplanOthers1996, Dourish2006}.
In 1996, Harrison and Dourish~\cite{HarrisonDourish1996} and in 2006 Dourish~\cite{Dourish2006} broadly discussed differences and similarities between the concepts of space and place and resulting implications for CSCW researchers.
An important outcome of their considerations, to which we relate our research, is the assumption that digital technologies, such as online sharing platforms, influence the way people encounter and appropriate urban space. 

We distinguish traditional understandings of (tourism) space as a defined geographical area from notions of (tourism) space as being socially constructed.
Traditional approaches are not satisfying anymore in order to understand current urban tourism phenomena.
Therefore, we motivate the constructionist approach we followed in our case study of two Berlin neighborhoods.
We argue that digital representations of space, as produced in Airbnb listing descriptions, have implications on the way tourists get to know about, encounter, and appropriate such areas.
Finally, we link these theoretical considerations to our research design and motivate our argument that urban tourism space is also digitally constructed, for example in the Airbnb listings that we chose to analyze empirically.  

\subsection{Traditional Understandings of Urban Tourism Space}

Traditionally, tourism space, in the sense of a tourist destination, has been defined as a geographical area~\cite{BurkartMedlik1976, DavidsonMaitland1997} that contains agglomerations of tourism-related attractions, facilities, and services~\cite{Pearce2013, SaraniemiKylanen2010}.
Accordingly, the destination is a stable and closed spatial unit, valorized for tourism purposes and filled up with tourism infrastructure.
It is the distinct physical space in which tourism, tourism development, politics, and planning happen.
In such traditional conceptualizations of tourism space~\cite{Framke2002}, the destination is understood as a static and given entity.
Natural and cultural resources, as well as tourism infrastructure, are its defining elements---tourism only happens within it.
These definitions entail a container-like notion of space.
As a result, the tourist destination is reduced to a confined space to which people are traveling for leisure reasons~\cite{Leiper1995}.

Urban tourism space has been framed similarly.
First illustrated in the \emph{tourist-historic city model}, designed by Ashworth and de Hahn in 1985, spatial clusters of historical sights, leisure facilities, food or accommodation infrastructure, were designated as tourism space~\cite{AshworthHaan1985}.
Like many other urban ecological or land use models that were commonly used in urban geography~\cite{Pacione2009}, the \emph{tourist-historic city model} divided the city into various functional regions---the tourism region being one of them. 
More specifically, a central section of the city, mostly the area intersecting between the historical center with its traditional buildings and the central business district (CBD) with all the shops, hotels, and restaurants, was considered to be tourism space~\cite{AshworthTunbridge1990, AshworthTunbridge2000}.
Such urban ecological models reflect scholars' intention to identify and separate the functional regions of a city, based on their prevalent built environment, that is the buildings, infrastructure, and other physical objects.
However, those models rarely capture dynamic aspects such as today's large flows of people, capital, and information~\cite{ShellerUrry2006}---they are static, but tourists' behavior is not.
The action space of tourists in the city has long been equated with a particular central urban region dedicated to them.
As a result of this spatial differentiation between tourism and non-tourism space, regions of the city lacking corresponding infrastructure cannot be touristically significant, following the inherent logic of the \emph{tourist-historic city model}~\cite{AshworthTunbridge1990, AshworthTunbridge2000}.

A spatial concentration of tourism in the city center was likely to be true at the time the \emph{tourist-historic city model} was designed.
Until well into the 1970s, tourism was not even regarded as a separate industry or function of the city~\cite{BurtenshawAshworthOthers1991}.
Cities were places for the serious tasks of work and government---tourism in contrast, was a function of the periphery~\cite{Christaller1964}.
People travelled from the city to the countryside for leisure reasons.
Consequently, researchers and urban governments did not perceive tourism as a field for intervention~\cite{Ashworth2003, AshworthTunbridge1990}.
They considered tourists to be invisible in all but a few selected districts of a few remarkable cities at specific times.
Their economic significance was peripheral~\cite{Ashworth2003} and tourism could hardly be isolated as a separate industry~\cite{AshworthTunbridge2000}.   
Only 50 years later, the perception of tourism has changed dramatically.

Urban tourism is booming in Western Europe since the 1990s~\cite{EurostatStatisticsExplained2016}.
This boom is both supply- and demand-side driven~\cite{Law2000, FreytagPopp2009}.
On the supply side, governmental urban regeneration schemes were carried out throughout Europe and North America, initiating the re-attractiveness of inner city areas.
Governmental promotion of tourism as a regeneration strategy started in the 1980s~\cite{Law1992}, when many cities suffered from de-industrialization, rising unemployment rates, and derelict manufacturing sites.
Urban tourism and the erection of respective infrastructure such as entertainment facilities~\cite{JuddFainstein1999, Spirou2010} were intended to initiate a counter-development in order to restructure the local economy, generating jobs and income~\cite{Law1992}.
At the same time, increasing affluence, more leisure time, and the ubiquitous proliferation of the car and the rise of the airplane as a means of transportation~\cite{Law1992} enabled people to travel not only to the countryside but also to urban destinations~\cite{FreytagPopp2009}.
After the fall of the Berlin Wall, Germany’s capital, which we focus on in our case study, became fully accessible for national and international tourists and visitors---their numbers exploded from 2003 onwards~\cite{AmtfurStatistikBerlinBrandenburg2017g}.  
Today's so-called \emph{``overtourism''}~\cite{Popp2012, PostmaSchmuecker2017} is becoming an emerging issue, not only in traditional tourist cities like Rome and Venice, but also in upcoming urban tourism destinations like Barcelona and Berlin.
As a result, residents are protesting against large tourist masses, flooding into their neighborhoods and \emph{``touristifying''} residential areas~\cite{ColombNovy2016, GravariBarbasGuinand2018, Gant2016}.


\subsection{Constructing New Urban Tourism Space}

Tourism in general and tourist behavior in particular have changed.
Tourists are no longer staying within the confines of the central tourist zone with major sights in walking distance and gastronomic infrastructure just around the corner.
Instead, many of them are now venturing into residential neighborhoods.
Urban tourism scholars discuss this phenomenon as \emph{off the beaten track} or \emph{new urban tourism}~\cite{MaitlandNewman2009, PappaleporeMaitlandOthers2010, PappaleporeMaitlandOthers2014, FullerMichel2014, StorsKagermeier2017}.
Reasons for those changing visitor practices are diverse.
One important factor is the rise of the mobile internet and the proliferation of information sharing platforms of the social web, such as Instagram, TripAdvisor, and the like.
These new information technologies are nowadays shaping the ways we encounter urban space~\cite{Dourish2006}.
They enable us to navigate through an unknown city~\cite{BrownPerry2001, Vertesi2008}, recommend us the best restaurant nearby~\cite{HicksCompOthers2012}, and let us stay in an Airbnb accommodation at our favorite location.
Thus, mobile technologies are no longer \emph{``simply operating within a spatial environment''}~\cite{BrewerDourish2008}.
Instead, they contribute to the \emph{``production of spatiality and spatial experiences''}~\cite{BrewerDourish2008}.
Their location-based service influence the way we move through the city and their personalized recommendation systems decide what we should see and where we should eat or sleep.
Consequently, mobile technologies direct our perception of a city and sometimes even suggest adequate place performances. 

Before the widespread usage of mobile technologies and online information sharing platforms, the travel guide and the city’s tourist information offices were the most influential information sources for visitors---and for many tourists they still are.
Travel guides contain descriptions of people and landscapes and pre-interpret them for visitors.
They influence peoples' decision-making process of what should be seen and thus affect visitors' perception of the city~\cite{Edensor1998}.
Urban management and marketing organizations play a similar role. 
They intend to attract and steer tourists by marking and marketing sights and places.
Iconic buildings are framed and promoted as must sees in live.
The Eiffel tower, for example, is signified as a symbol for love, it is endowed with this particular meaning.  
Gazing upon the sight thus turns into a romantic experience for many. 
This example illustrates that iconic buildings, like places, have no \emph{``intrinsic attraction power''}~\cite{Gunn1972, Leiper1990}.
Instead, they are formed and fashioned by human beings~\cite{Iwashita2003}.
They are marked and signified as iconic architecture worth visiting~\cite{MacCannell1976, Saarinen2004, Edensor1998}, meanings and values are ascribed to them~\cite{Squire1994}, and they are enacted through social practice~\cite{Edensor2000, Edensor2001, Larsen2008, Edensor1998, BrenholdtHaldrupOthers2004}.

In many cities around the world, the DMOs are responsible for representing their city.
In a process of \emph{``place branding'}'~\cite{MoilanenRainisto2009}, they develop a strategy to promote the urban destination, utilizing tools such as \emph{``policy making, planning, advertisement campaigns, exhibitions, publicity, and the like''}~\cite{ChenChen2016, Geary2013}.
As a result, urban representatives create images about cities and places purposefully and strategically in order to succeed in a fierce competition for visitors.
Such space images, manifested in flyers, maps, pictures---both materially and digitally---are fragments of the whole discursive framework co-constructing the city ~\cite{Saarinen2004}. 
As several researchers have pointed out, spatial images contributing to a discourse are never neutral.
(Spatial) discourses are socially \emph{``produced coherent meaning systems and practices, which both manifest and are power structures at the same time''}~\cite{Saarinen2004}.
Power geometries~\cite{Dourish2006} and ideologies are inherent in spatial discourses~\cite{Lefebvre1991}.
Consequently, the destination image produced by a city's DMO is a manifestation of their power.
They have the ability to legitimize one space image over the others~\cite{Davis2005}.

Due to the rise of mobile technologies and several peer-to-peer (information) sharing platforms, such as Yelp, TripAdvisor, and Airbnb, we argue that the above-mentioned power structures are shifting.
Following Dourish’s point of view~\cite{Dourish2006}, we establish that such technologies influence the way people move through the city and encounter urban space.
Moreover, information on online platforms is no longer solely provided by the city's DMO.
Due to social web applications, many people can now produce content online, for example to market their local business or their Airbnb apartment.
In doing so, they produce their own spatial representations of the city or neighborhood and thus participate in the spatial discourse.
Digital technologies in general and Airbnb listings in particular empower local people to participate in the digital co-production of urban (tourism) space. 

For urban visitors, mobile technologies open up spaces they would hardly have encountered before.
Classic tourist maps, for example, only represent a fraction of the city and visitors tend to stay within this confined area.
Digital peer-to-peer platforms, in contrast, also provide information on sites or places outside the central tourist district(s).
Thus, visitors get to know about rather unexplored parts of the city and are motivated to leave the central tourist area.
Mobile technologies enable visitors to access on-site information about sights and places and help them to navigate their way through the city.

Moreover, a rising number of tourists are searching for authentic experiences \emph{off the beaten track} and want to immerse themselves in the local, everyday life~\cite{PappaleporeMaitlandOthers2014}.
This perspective challenges traditional conceptualizations of tourism as an escape from work and home~\cite{Urry1990}.
Traditional understandings of urban tourism, as represented in the \emph{tourist-historic city model}, associate residential neighborhoods with mundane activities.
Tourism, in contrast, has long been regarded as an extraordinary practice that happens at distinct times and places, separating people \emph{``off from everyday experiences''}~\cite{Urry1990}.
From this point of view, tourism is unlikely to happen in mundane residential areas.
Actual visitor practices challenge such traditional binary differentiations between work-leisure, tourist-resident, and home-away~\cite{CohenCohen2017}.
Those conceptual boundaries become increasingly blurred when visitors venture into residential neighborhoods, stay in private Airbnb apartments, and behave like locals.
\emph{Off the beaten track} tourists intend to differentiate themselves spatially and ideationally from mass tourism, which is still located in the central areas of the city~\cite{Freytag2010, McCabe2005}.
Against this background, the \emph{tourist-historic city model} is losing its explanatory power.
Tourism is no longer restricted to certain areas, but can take place all over the city.

\begin{figure*}
\centering
\includegraphics[width=0.8\columnwidth,  trim=0.0in 0.0in 0.0in 0.0in]{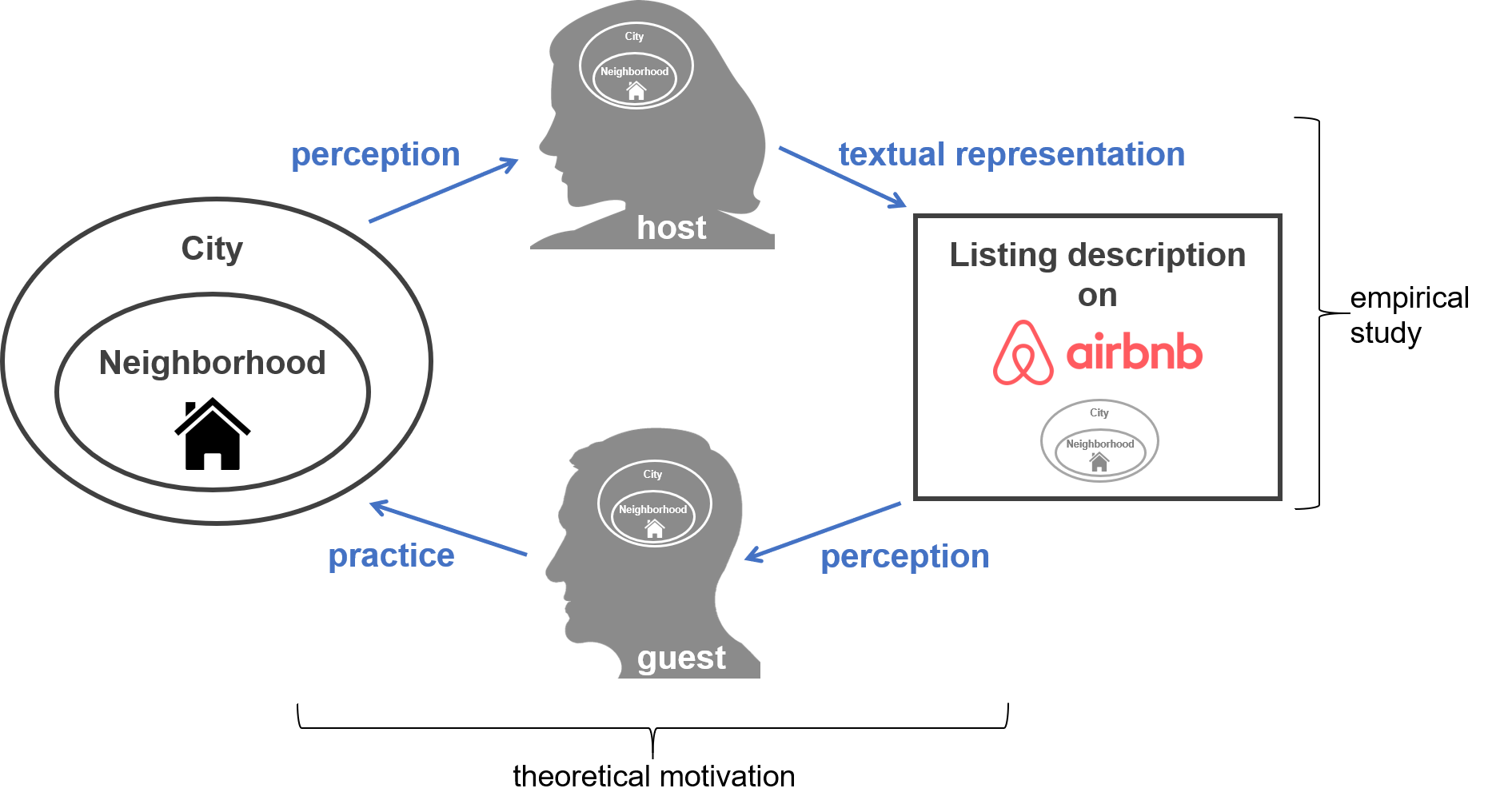} 
\caption{Overview of our conceptual approach: Airbnb hosts perceive their city and neighborhood and capture this perception in their listing descriptions. Guests read those descriptions, which contribute to their image of the neighborhood they are going to visit. This, again, influences guests' practices during their visit. As we focus on how Airbnb listing descriptions may influence guests' practice, we don't show other relationships such as guests' perception of neighborhood and city and hosts' practice in that environment.}
\label{fig:overview}
\end{figure*}

Most research on \emph{off the beaten track} and \emph{new urban tourism} has focused on the tourist perspective and their desires to leave the central tourist zones.
What is still lacking in scholarly research is the spatial perspective and attempts to explain how residential neighborhoods become touristically significant.
The intention of our research is to tackle this gap and, in particular, to analyze how residential neighborhoods are transformed into tourist places.
To this end, we follow a constructionist understanding of urban tourism space.
That means we no longer regard tourism facilities and infrastructure as the central elements defining tourism space.
Instead, we assume that tourism space is socially constructed through representations and performances, as illustrated in Figure~\ref{fig:overview}.
For us, people are the major agents who transform places and landscapes into tourism destinations.
They have the ability to transform space both physically and materially as well as perceptually and symbolically~\cite{Iwashita2003}.
This production of space, however, does not only happen in the physical realm.
Following some initial research approaches in geography~\cite{ZukinLindemanOthers2015} and CSCW~\cite{Corbett2017}, we analyze representations of space that are produced digitally and collaboratively.
Contrary to most research looking into space images produced by the DMO~\cite{ChenChen2016, PritchardMorgan2000}, we take spatial representations into account that are produced by local residents in their Airbnb listing descriptions.
We argue that through the means of digital technologies, local people are now empowered to participate in the discourse producing and reproducing urban space.

\subsection{From Theoretical Concepts to Empirical Research}

The city depiction in Figure~\ref{fig:overview} represents the city in its multitude of dimensions.
It stands for buildings and infrastructure, but likewise for people, for flows of information and capital, for images and feelings associated with it, for its various representations in pictures and discourses, and for everyday practices enacting it.
The Airbnb hosts in our case study can never grasp the full urban system, they only perceive a part of reality, that is their imagination of the city, existing in their minds.
This imagination is never fixed, it is subject to numerous external and internal factors, and can easily change.

However, when hosts decide to rent out a room or an apartment on Airbnb, the platform’s listing structure encourages them to write down information about the neighborhood wherein the apartment is located.
In the process of writing, hosts reproduce their space image~\cite{BalogluMcCleary1999, KockJosiassenOthers2016}.
Here again, the listing description is not a full duplicate or total representation of everything hosts know and feel about the city and the neighborhood---it is a selection of aspects that they find interesting or important to know.
At the same time, the listing description is written for a certain audience.
Since hosts intend to rent out their room or apartment, the text has a promotional character.
Hosts deliberately produce a particular space image~\cite{ChenChen2016, Young1999}. 
The sum of neighborhood descriptions posted on Airbnb finally results in a collaboratively produced digital representation of (imagined) urban space.
It thus contributes to the discursive framework of the respective area.

The proliferation of (information) sharing platforms, such as Instagram, TripAdvisor, and Airbnb, enables people to signify places or buildings to be worth visiting.
For example, remote rocks in Norway are being reproduced thousands of times on Instagram and thus become a tourist attraction~\cite{Meier2018}.
These platforms empower people to reinterpret places, to ascribe new meanings to them, and hence to transform them into places of significance for visitors.
Finally, in producing such online or offline, written or filmed, pictorial or oral representations of space, every participant contributes to the way space is constructed and perceived~\cite{Saarinen2004}.

After posting the neighborhood description on Airbnb, these digital textual representations of urban space are read and processed by potential guests, affecting their space image.
However, Airbnb listings are not the only source of information that guests use.
For the most part, they already have space images in their minds, which are based on previous visits, newspaper articles, tourist guidebooks, tips from friends and relatives, or posts on Instagram or TripAdvisor.
All these sources of information influence, to a certain extent, what guests are going to do during their stay~\cite{Larsen2012}.

In our research, we decided to focus on Airbnb listing descriptions of the neighborhood, because they have a high potential to steer visitors’ attention, in particular to elements that would otherwise remain unrecognized.
Against the background of \emph{off the beaten track} and \emph{new urban tourism}, scholars have already illustrated that visitors want to behave like a local~\cite{PappaleporeMaitlandOthers2010, PappaleporeMaitlandOthers2014}.
In order to do so, they rely on insider tips that hosts can provide. 
Staying in a private Airbnb room or apartment enables hosts and guests to exchange expectations and experiences.
Guests receive local information, for example about the best restaurant or coolest club, and get to know local routines and practices.
Instead of discussing  spots for picture taking and gazing~\cite{Urry1990} upon sights, hosts rather provide information on the closest route for a morning jog.
Such information allows guests to take part in the local everyday life~\cite{Larsen2008, Larsen2012}.
As a result, listing descriptions on Airbnb do not only produce space images and influence the way in which space is perceived, but also impact visitors' behavior in place.
This is because Airbnb hosts encourage guests to visit certain places and motivate related practices (see Section~\ref{sec:results}). 

In our case study, we empirically analyze how hosts describe and thus co-produce the image of their neighborhoods.
We have theoretically motivated the relation of those digital descriptions to the physical world, that is the neighborhood and the city, and peoples' performances in that environment. 
We consider it to be an important direction for future work to also analyze Airbnb from a user perspective.
On the one hand, one could evaluate how hosts write and revise their neighborhood descriptions.
On the other hand, it would be interesting to see how exactly visitors read those descriptions and how they influence their practices.

\section{Research Design}

\begin{figure*}
\centering
\includegraphics[width=1\columnwidth,  trim=0.0in 0.0in 0.0in 0.0in]{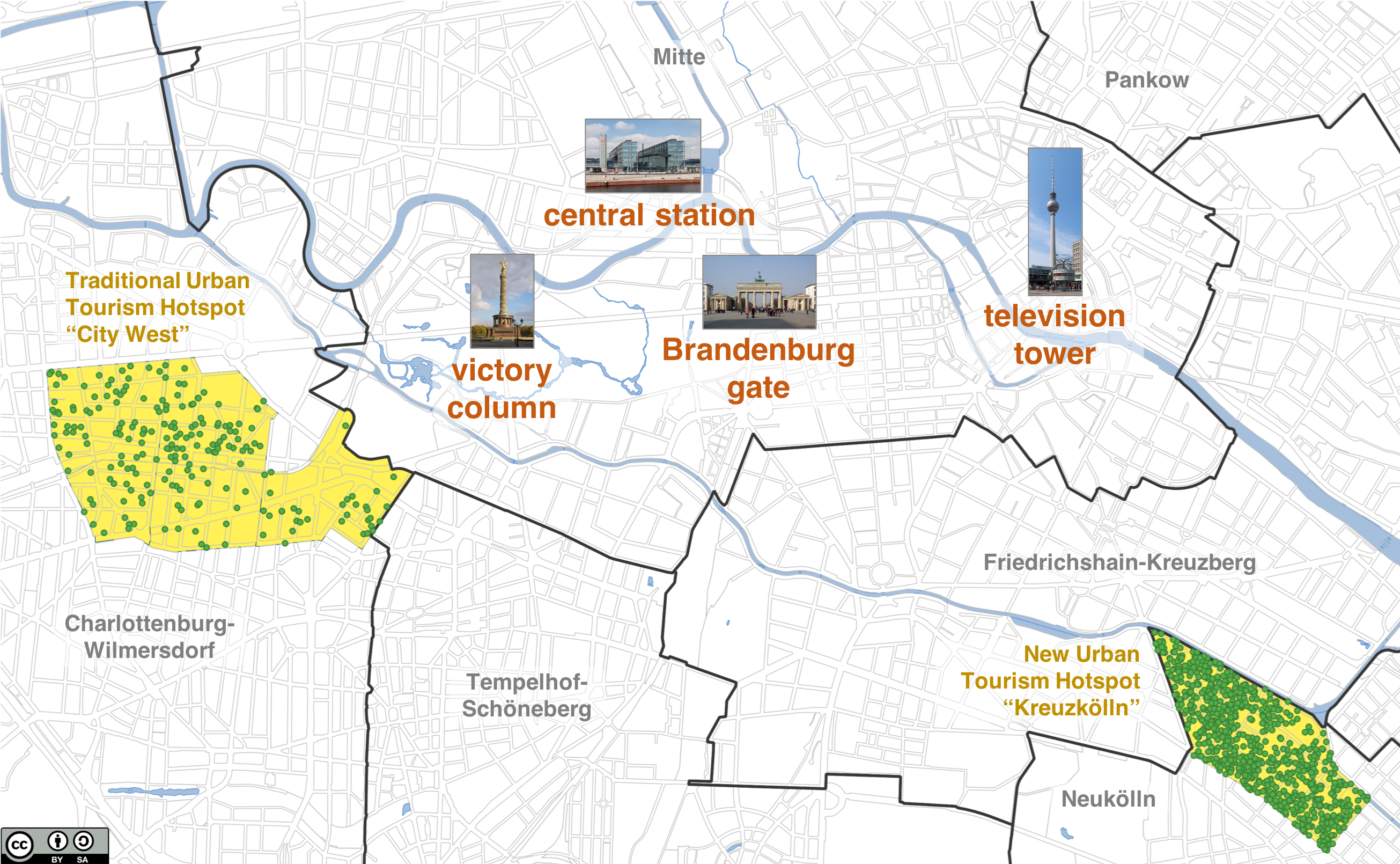} 
\caption{Analyzed areas (yellow) and identified Airbnb listings (green, $n=965$); map data from Amt für Statistik Berlin-Brandenburg~\cite{AmtfurStatistikBerlinBrandenburg2016, AmtfurStatistikBerlinBrandenburg2017b, AmtfurStatistikBerlinBrandenburg2017c, SenatsverwaltungfurStadtentwicklungundWohnenBerlin2018}, Airbnb listings retrieved 2017-03-22 (see Section~\ref{sec:data-collection}), images from Wikipedia.}
\label{fig:erhebungsraum}
\end{figure*}

As motivated above, our goal was to analyze and compare how the two neighborhoods \emph{Kreuzkölln} and \emph{City West} are constructed as tourism spaces in Airbnb listing descriptions (RQ1 and RQ2).
For our analysis, we deliberately chose two very different neighborhoods.
The neighborhood we denote \emph{City West} has a long tradition of being touristically significant.
It contains internationally known iconic sights~\cite{visitBerlin2018f, visitBerlin2018g} that are visited by a large number of people every year.
From January until July 2017, the borough \emph{Charlottenburg-Wilmersdorf} had about 1.5 million overnight guests~\cite{AmtfurStatistikBerlinBrandenburg2017f}.
Against this background, we consider \emph{City West} to be a \emph{traditional urban tourism hotspot}. 
The neighborhood we denote as \emph{Kreuzkölln}, in contrast, had a rather difficult past.
It has long been associated with poverty, crime, and drugs~\cite{AmtfurStatistikBerlinBrandenburg2017d, AbgeordnetenhausBerlin2017}, although its image is changing now~\cite{DeutschePresseAgentur2018}.
The neighborhood itself is lacking major sights, but the borough provides some facilities that are being promoted by Berlin's DMO~\cite{visitBerlin2018b}.
Only about 219,000 officially registered overnight guests stayed in the borough of \emph{Neukölln} between January and July 2017~\cite{AmtfurStatistikBerlinBrandenburg2017f}.
Despite these facts, \emph{Kreuzkölln} increasingly attracts tourists~\cite{SchultePeevers2017, FullerMichel2014}, which is why we consider the neighborhood to be a \emph{new urban tourism hotspot}.   

Our \emph{units of observation} are the free text fields of the Airbnb listings located in those neighborhoods.
This includes the listings' title, description, and house rules (see Figure~\ref{fig:airbnb-listing}).
Our \emph{units of analysis} are the hosts' descriptions of the city and the neighborhoods (outside)---the descriptions of the apartments (inside) are not in the focus of our research.
To delineate the two neighborhoods, we rely on the so-called \emph{Lebensweltlich Orientierte Räume} (LOR).
These urban planning areas have been introduced by the Berlin administration for urban development and housing in the year 2006.
Instead of focusing solely on infrastructure, the LORs also consider other factors such as social milieus and population density~\cite{SenatsverwaltungfurStadtentwicklungundWohnen2009}.


The neighborhood that we denote \emph{Kreuzkölln} corresponds to the LOR \emph{Reuterkiez} with 27,792 inhabitants (as of December 2016~\cite{AmtfurStatistikBerlinBrandenburg2017}).
As the density of Airbnb listings is smaller in the \emph{City West} neighborhood, we decided to include several LORs around the major West Berlin streets \emph{Kantstraße} and \emph{Kurfürstendamm}.
The LORs we selected are named \emph{Karl-August-Platz}, \emph{Savignyplatz}, \emph{Hindemithplatz}, \emph{Georg-Grosz-Platz}, and \emph{Breitscheidplatz}.
In total, 37,225 inhabitants live in those LORs (as of December 2016~\cite{AmtfurStatistikBerlinBrandenburg2017}).
Figure~\ref{fig:erhebungsraum} visualizes the location of the two areas. We provide the retrieved host and listing data, the R scripts used for our quantitative analysis, and our final coding schema as supplementary material~\cite{StorsBaltes2018}.

The last step of our research was to compare hosts' neighborhood descriptions with the descriptions of \emph{visitBerlin}, the city's DMO (RQ3).
To this end, we extracted all places that \emph{visitBerlin} mentions in the corresponding borough descriptions on their website~\cite{visitBerlin2018h, visitBerlin2018i}.
Then, we compared those places with the most frequently named places in the Airbnb listings.
Since we identified a disparity between what Airbnb hosts and the DMO consider important places, we further analyzed this relationship.
We assessed how frequently the places named in the DMO's borough descriptions are mentioned in the Airbnb listing descriptions we retrieved for the two neighborhoods (see Sections~\ref{sec:data-collection} and \ref{sec:quantitative-analysis}).

\subsection{Data Collection}
\label{sec:data-collection}

To retrieve the Airbnb listings in the two neighborhoods, we utilized Tom Slee's \emph{airbnb-data-collection} tool\footnote{\url{https://github.com/tomslee/airbnb-data-collection}}.
On March 22, 2017, we collected all listings within two bounding boxes encompassing the \emph{Kreuzkölln} and \emph{City West} areas.
Afterwards, we utilized the LOR shapefiles provided by the city of Berlin~\cite{AmtfurStatistikBerlinBrandenburg2016} to filter out listings that are not located within the selected LORs.
Please note that Airbnb slightly randomizes the location of listings.
Consequently, some listings at the border of the neighborhoods may actually be located in the surrounding LORs (or vice versa).
As the neighboring LORs are very similar to the analyzed ones in terms of their social structure and urban fabric~\cite{SenatsverwaltungfurGesundheitPflegeundGleichstellungBerlin2014, Levy1999}, this should not pose a threat to the validity of our results.
In the end, we were able to identify 753 listings in the \emph{Kreuzkölln} area and 212 listings in the \emph{City West} area.
This means that there were five times more listings per inhabitant in the new urban tourism hotspot \emph{Kreuzkölln} (0.03 listings per inhabitant) compared to the traditional urban tourism hotspot \emph{City West} (0.006 listings per inhabitant).
In other words, statistically, 1 out of 33 inhabitants in \emph{Kreuzkölln} rents out a room or an apartment on Airbnb---opposed to 1 out of 167 in \emph{City West}.


To be able to quantitatively describe the retrieved listings and to analyze their descriptions, we needed to retrieve additional information, which was not possible with the above-mentioned tool.
Thus, we developed our own tool, which utilizes Airbnb's unofficial API ~\cite{Baltes2017c}.
Using this tool, we were able to successfully retrieve data for 750 listings in the \emph{Kreuzkölln} neighborhood and for 210 listings in the \emph{City West} neighborhood.
The retrieval failed for five listings.
In the following, we briefly describe the hosts and listings in the neighborhoods, before we continue with our research design.

The daily price of the analyzed listings differed significantly in the two neighborhoods:
The average daily price in \emph{Kreuzkölln} was 51.65 Euro ($Mdn=45$, $SD=37.84$) compared to 73.71 Euro in \emph{City West} ($Mdn=55$, $SD=64.88$).
The difference was significant according to the nonparametric two-sided \textit{Wilcoxon rank-sum test}~\cite{Wilcoxon1945} with p-value $<\num{8.0e-9}$.
However, the effect was only small according to \textit{Cohen's} $d$~\cite{Cohen1988, GibbonsHedekerOthers1993}, considering the thresholds described by Cohen~\cite{Cohen1992} ($d=0.49$).
A factor that could bias the listing price is the accommodation type (whole apartments are usually more expensive than single rooms).
However, the distribution of listing types was quite similar in both neighborhoods: In \emph{Kreuzkölln}, there were 390 (52.0\%) entire homes or apartments, 358 (47.7\%) private rooms, and 2 (0.3\%) shared rooms.
In \emph{City West}, there were 116 (55.2\%) entire homes or apartments, 90 (42.9\%) private rooms, and 4 (1.9\%) shared rooms.

\begin{table}
\small
\centering
\caption{Professionalism of hosts in the two neighborhoods, measured by the total number of listings they provide on Airbnb (the listings are not necessarily located in one of the two neighborhoods, because hosts may also provide listings in other areas of Berlin or in other cities worldwide).}
\label{tab:listings-per-host}
\begin{tabular}{c | rrrrr r}
\hline
\multicolumn{1}{c|}{Listing created by host} & \multicolumn{1}{c}{\multirow{2}{*}{1}} & \multicolumn{1}{c}{\multirow{2}{*}{2}} & \multicolumn{1}{c}{\multirow{2}{*}{3}} & \multicolumn{1}{c}{\multirow{2}{*}{>3}} & \multicolumn{1}{c}{\multirow{2}{*}{NA}} & \multicolumn{1}{|c}{\multirow{2}{*}{...listings}}  \\
\multicolumn{1}{c|}{with a total of...} & & & & & & \multicolumn{1}{|c}{} \\
\hline
\hline
Kreuzkölln & 622 (83.2\%) & 91 (12.2\%) & 20 (2.7\%) & 15 (2.0\%) & 2 (0.3\%) & \multicolumn{1}{|r}{750 (100\%)} \\
City West & 144 (68.9\%) & 39 (18.7\%) & 9 (4.3\%) & 17 (8.1\%) & 1 (0.7\%) & \multicolumn{1}{|r}{210 (100\%)} \\
\hline
\hline
Total & 766 (80.0\%) & 130 (13.6\%) & 29 (3.0\%) & 32 (3.3\%) & 3 (0.4\%) & \multicolumn{1}{|r}{960 (100\%)} \\
\hline
\end{tabular}
\end{table}

The listings were provided by 898 different hosts (708 in \emph{Kreuzkölln} and 190 in \emph{City West}).
To assess the number of professional hosts in the two areas, we retrieved the total number of Airbnb listings that those hosts provide.
This includes listings in other areas of Berlin or other cities worldwide.
We were able to successfully retrieve this information for 706 hosts in \emph{Kreuzkölln} and for 189 hosts in \emph{City West}.
The retrieval failed for three hosts.
There is no universally accepted threshold for the number of listings that a host must provide to be considered \emph{professional}.
Nevertheless, we observed that most hosts (80\%) in the two areas provide only one listing.
However, there is a considerable difference between the two neighborhoods: 31.1\% of the listings in \emph{City West} were provided by hosts with more than one listing, opposed to 16.9\% in \emph{Kreuzkölln} (see Table~\ref{tab:listings-per-host}).
The degree of professional hosting on Airbnb seems to be higher in the traditional tourist hotspot \emph{City West} than in the new urban tourism hotspot \emph{Kreuzkölln}.


To compare hosts' descriptions in the listings to \emph{visitBerlin}'s descriptions on their website (RQ3), we retrieved the English version of the borough pages of \emph{Neukölln}~\cite{visitBerlin2018h} and \emph{Charlottenburg-Wilmersdorf}~\cite{visitBerlin2018i} on July 5, 2018.
Afterwards, we manually extracted all specific place names from those pages, ignoring vague descriptions.
In the description of the borough \emph{Neukölln}, for example, the DMO refers to the borough's diverse built environment \emph{``from its estates of detached houses in the south to the high-rises in the Gropiusstadt neighbourhood''}~\cite{visitBerlin2018h}.
Here, we only considered the specific place \emph{Gropiusstadt neighbourhood} and ignored the vague description of \emph{estates of detached houses in the south}.
Both borough pages, \emph{Neukölln} and \emph{Charlottenburg-Wilmersdorf}, have the same structure:

\begin{itemize}
\item A header with the borough's name, including a slogan describing the area,
\item pictures and brief descriptions of the DMO's \emph{``favorite places''} in the borough,
\item a section about \emph{``what you need to know''} about the borough,
\item and several paragraphs on selected topics.
\end{itemize}

Table~\ref{tab:places-dmo} lists the places mentioned by the DMO together with the number and percentage of Airbnb listings in which those places were likewise mentioned (see Section~\ref{sec:quantitative-analysis}).
In their borough descriptions, \emph{visitBerlin} sometimes used places only to describe the location of other places or facilities.
In the description of \emph{Neukölln}, for example, the DMO mentions \emph{``artist studios located between the Landwehr Canal, Sonnenallee and Hermannstraße}''~\cite{visitBerlin2018h}.
Here, the artist studios are in focus, not the canal and the two streets, which are merely used to describe the location of those studios.
In the table, we marked such places with an asterisk and used a lighter background color.

\subsection{Qualitative Data Analysis}
\label{sec:data-analysis}

\begin{figure*}
\centering
\includegraphics[width=\columnwidth,  trim=0.0in 0.0in 0.0in 0.0in]{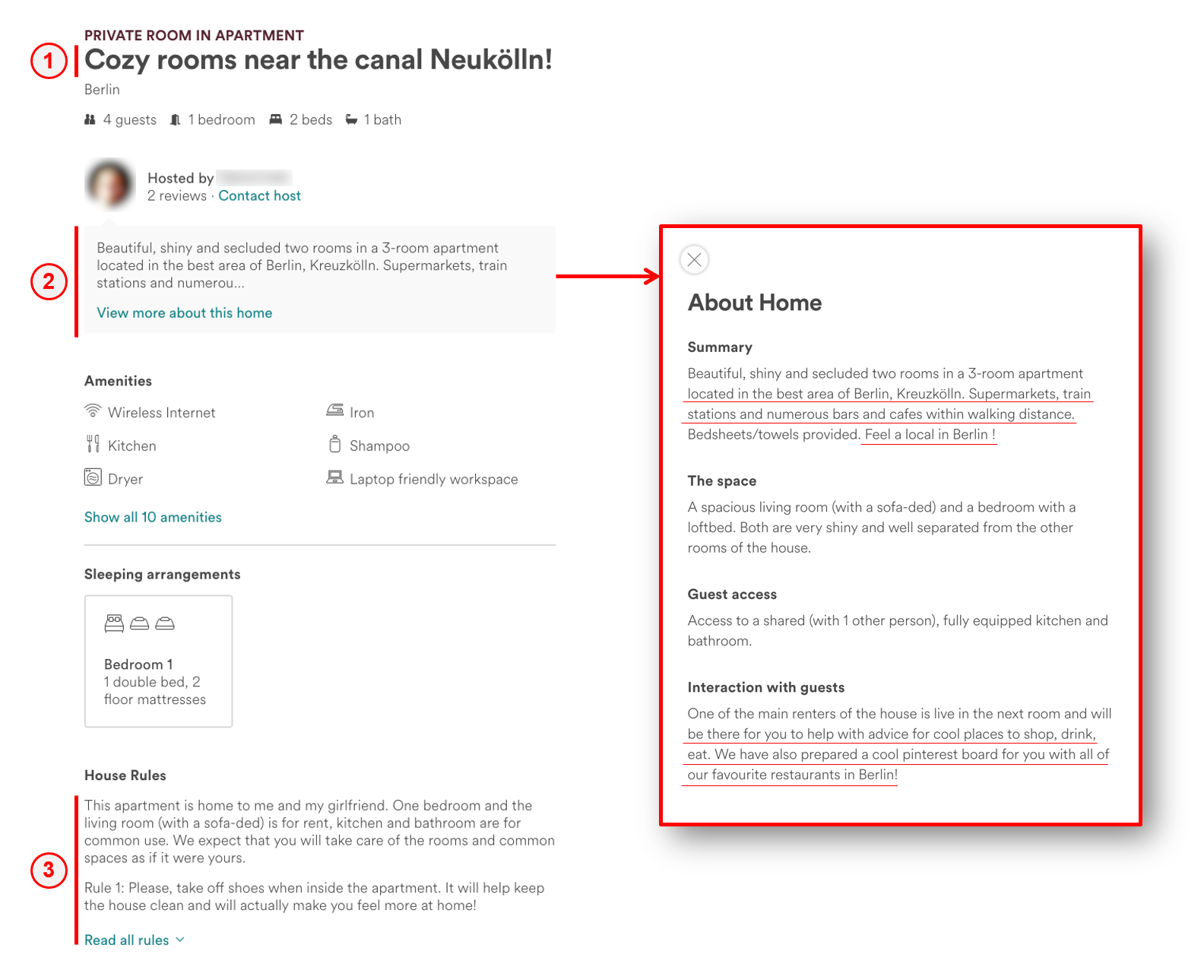} 
\caption{Exemplary Airbnb listing in the Kreuzkölln neighborhood (\url{https://airbnb.com/rooms/7738887}). For all listings in the samples, we qualitatively analyzed the title (1), the complete listing description (2), and the house rules (3).
}
\label{fig:airbnb-listing}
\end{figure*}

Figure~\ref{fig:airbnb-listing} shows an exemplary Airbnb listing from the \emph{Kreuzkölln} neighborhood and visualizes where the information we analyzed is located on the Airbnb website.
The median length of the listing descriptions (including house rules) was 138.5 words for \emph{Kreuzkölln} and 117.5 words for \emph{City West}; the median title length was five words for both neighborhoods.
To qualitatively analyze and compare the data, we drew a random sample of 100 listings (50 from each neighborhood) and imported them into the CAQDAS software \emph{MAXQDA}.\footnote{\url{https://www.maxqda.com/}}
The subsequent qualitative analysis consisted of three phases: First, the two authors coded 25 listings from each neighborhood independently (\emph{initial coding}~\cite{Saldana2015}).
In a second phase, the authors discussed their codings until they agreed on a common coding schema.
In the last phase, one author coded the remaining 50 listings using the common coding schema (\emph{elaborative coding}~\cite{Saldana2015}) and after that, both authors discussed the final coding.
The final coding schema focused on four main topics:

\begin{enumerate}
\item Which \emph{places} in the neighborhoods and in the city are mentioned?
\item Which \emph{facilities} and \emph{people} in the neighborhoods are described?
\item Which \emph{adjectives} are used to describe the neighborhoods?
\item Which \emph{practices} are encouraged in the descriptions?
\end{enumerate}

For aspects (1), (2), and (3), we conducted a \emph{word-by-word coding}~\cite{Charmaz2014}, assigning the places / facilities / adjectives to corresponding categories that emerged from the data.
For aspect (4), an \emph{in-vivo coding}~\cite{Charmaz2014, Saldana2015} approach was more appropriate.
We assigned whole sentences encouraging certain practices to categories that again emerged from the data.
In one \emph{Kreuzkölln} listing, for example, a host wrote the following: \emph{``Club amateur, let's enter the About Blank, Berghain or Griessmühl by bike! You'll look like proper Berliner! We are located from each for less than 15 minutes.''}
We assigned this statement to the practice category \textsc{go clubbing/enjoy nightlife}.

\subsection{Quantitative Data Analysis}
\label{sec:quantitative-analysis}

After we finished our qualitative analysis of the listing descriptions, we used the resulting coding schema to compare those descriptions to the borough descriptions of \emph{visitBerlin} (RQ3).
We extracted the places from the borough descriptions on the DMO's website, as described in Section~\ref{sec:data-collection}, and compared those places with the places frequently mentioned by Airbnb hosts.
Since we identified a disparity between what Airbnb hosts and the DMO consider important places (see Section~\ref{sec:comparison-dmo}), we decided to quantitatively search for the places mentioned by the DMO in their descriptions of \emph{Neukölln} and \emph{Charlottenburg-Wilmersdorf} in all 960 Airbnb listing descriptions we retrieved for the two neighborhoods \emph{Kreuzkölln} (located in \emph{Neukölln}) and \emph{City West} (located in \emph{Charlottenburg-Wilmersdorf}).
That way, we were able to estimate the overlap between places mentioned by the hosts and the DMO.

For each extracted place, we built a regular expression matching different spellings of the place.
The `Kurfürstendamm', for example, is also known as `Ku'damm' or `Kudamm'.
Moreover, the German umlaut `ü' can be represented as `ue', thus `Kurfuerstendamm' is an additional possible spelling.
The regular expression we used in that case was:

\begin{quote}
\begin{verbatim}
(?i:.*ku[^\n\t]+damm[^\n\t]*)
\end{verbatim}
\end{quote}

The regular expression is case-insensitive and matches the complete line in which the pattern was found.
Since the Kudamm was mentioned in the DMO's description of \emph{Charlottenburg-Wilmersdorf}, where the \emph{City West} neighborhood is located, we only searched
the Airbnb listing descriptions we retrieved for that neighborhood.
We utilized the programming language \emph{R} to search the listing descriptions for matches of the regular expression and found matches in 119 of 210 descriptions (56.7\%).
We used this workflow for all places mentioned by the DMO and provide the R script containing the regular expressions as supplementary material~\cite{StorsBaltes2018}.
The result of this analysis can be found in Table~\ref{tab:places-dmo}.
Section~\ref{sec:comparison-dmo} summarizes our findings.

\section{Results}
\label{sec:results}

\begin{figure*}
\centering
\includegraphics[width=0.75\columnwidth,  trim=0.0in 0.0in 0.0in 0.0in]{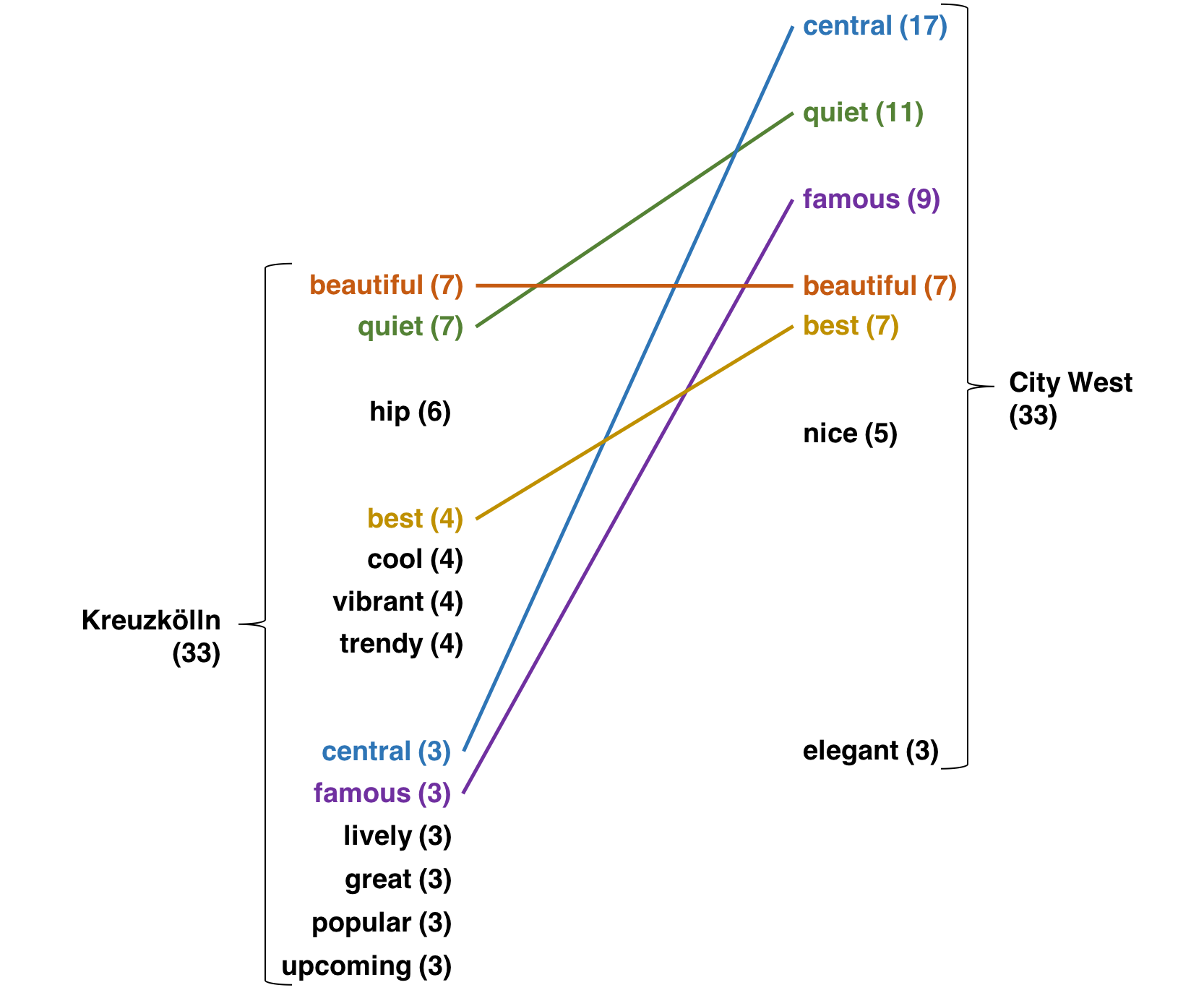} 
\caption{Adjectives used to describe the neighborhood (only adjectives mentioned in $\ge 3$ listings are shown).}
\label{fig:adjectives}
\end{figure*}

In this section, we summarize key findings from our qualitative and quantitative data analyses. 
As described above, the four high-level concepts \emph{places}, \emph{facilities}, \emph{adjectives}, and \emph{practices} emerged from the qualitative data.
Table~\ref{tab:places} shows the number of listings in which particular \emph{places} in the neighborhood or other parts of the city were mentioned.
Table~\ref{tab:facilities} shows how many listings contained information on \emph{facilities} in the vicinity of the offered apartment or room.
An important aspect for the construction of the neighborhoods are the \emph{adjectives} that hosts use in their descriptions.
In our analysis, we only considered adjectives used to describe streets, places, or the whole neighborhood, excluding descriptions of the apartment or room.
Figure~\ref{fig:adjectives} shows the adjectives that were mentioned in at least three different listings in one of the neighborhoods.
We ordered them according to the number of listings they were used in and highlighted the ones that were used in the descriptions of both neighborhoods.
The encouraged \emph{practices} are described throughout this section.
In the following, we focus on the aspects that are most suitable to illustrate differences between the two neighborhoods.
We provide the listing ID when referring to specific listings and provide the number of coded listings for important codes or categories.
To distinguish those two attributes, the number of coded listings is in \textbf{bold} font and the listing ID is in regular font.

\subsection{Places and Facilities}
\label{sec:places-facilities}

To spatially reference the location of an Airbnb room or apartment in the city, it was common for hosts to name specific places in the neighborhood or in the city.
This is particularly true for the short listing headlines, which we analyzed first.
The most frequently mentioned places in those headlines were \emph{Berlin}, \emph{Neukölln}, and \emph{Kurfürstendamm} (often abbreviated as \emph{Kudamm}): 

\begin{quote}
\emph{``Stay in the middle of West Berlin''} (10103432)\\
\emph{``Sunny room in the Heart of Neukölln''} (11321524)\\
\emph{``Bright DREAM LOFT at KuDamm 6 rooms''} (6931794)
\end{quote}

For visitors, the location of their accommodation is important.
Being located centrally or peripherally in the city affects the time needed to get around.
Moreover, an accommodation in close proximity to the facilities one intends to visit is often considered to be more comfortable.
However, the Airbnb hosts in our case study did not promote their apartment through naming distinct sights in their listing headlines.
Instead, they focused on places.
As headlines lack additional context, potential guests need to carry certain space images in mind in order to understand what these places are about and what meanings they convey.
Visitors have to decode the information inherent in such place names, particularly when solely reading the short listing headlines.   

Full listing descriptions, in contrast, contain much more information about certain places.
In the case of \emph{Kurfürstendamm}, for example, hosts mainly referred to the large variety of shopping facilities along the street: \emph{``Kurfürstendamm, the 3.5 kilometer shopping paradise''} (65340).
Naming \emph{Neukölln} in a listing description may produce a very different space image.

The borough \emph{Neukölln} has long been associated with poverty, crime, and drugs.
To this day, Neukölln is still the borough with the highest share of inhabitants threatened by poverty---in 2016, more than a quarter of its residents received social welfare~\cite{AmtfurStatistikBerlinBrandenburg2017d}.
The borough has a very high share of residents with migration background (46\% as of June 2017~\cite{AmtfurStatistikBerlinBrandenburg2017e}) and drugs are still a prevalent issue~\cite{AbgeordnetenhausBerlin2017}.
Nevertheless, the spatial reference \emph{Neukölln} \textbf{(41)} was named almost as frequently as \emph{Berlin} \textbf{(49)} over all analyzed Airbnb-listings.
80\% of the listings located in the neighborhood \emph{Kreuzkölln} referred to the borough's name.
For hosts, it seems to be of particular relevance that their rooms or apartments are located in \emph{Neukölln}.
Also the neighborhood's informal name \emph{Kreuzkölln}, which we used to denote the research area, was mentioned several times \textbf{(15)}.
It combines the names of the two neighboring districts \emph{Kreuzberg} \textbf{(15)} and \emph{Neukölln}, but at the same time, it represents much more.
The name \emph{Kreuzkölln} conveys a certain neighborhood image, including elements of the neighborhood's past, its recent development, typical facilities, and people who are attracted by its special mix:

\begin{quote}
\emph{``Kreuzkölln nightlife has almost outstripped Kreuzberg. New galleries, bars and restaurants are opening almost every week''} (178347)\\
\emph{``(...) the pulsing heart of Kreuzkölln district (one of the most hip and fastest developing areas since 2008)''} (912787)\\
\emph{``To be clear, this area is called the “Brooklyn of Berlin”. Now you get the picture! Neukölln is one of the most appreciated parts of the German capital, the epicenter of hip''} (1559030)
\end{quote}

Airbnb hosts write their listing descriptions with a certain intention, namely to promote their offered room or apartment.
As illustrated above, the name \emph{Neukölln} no longer solely refers to the social difficulties that the borough's government is facing.
After a decade of urban renewal initiatives and still on-going projects~\cite{SenatsverwaltungfurStadtentwicklungundWohnenBerlin2007, SenatsverwaltungfurStadtentwicklungundWohnenBerlin2011}, the neighborhood’s image is shifting.
Airbnb hosts pick up and highlight this development in order to promote their rooms and apartments for visitors.
In doing so, they contribute to the public discourse~\cite{DeutschePresseAgentur2017, Connolly2016, Dyckhoff2011, DeutschePresseAgentur2018}, framing the neighborhood as a \emph{trendy} and \emph{upcoming} area, a so-called \emph{Szenekiez}.
The adjectives illustrated in Figure~\ref{fig:adjectives} directly refer to the neighborhood's transformation (\emph{upcoming}) and its new image (\emph{hip}, \emph{cool}, \emph{vibrant}, and \emph{trendy}). 

The district \emph{Charlottenburg} \textbf{(21)}, which belongs to the larger borough \emph{Charlotten\-burg\--\-Wilmers\-dorf}, is described very differently.
The district's name is not mentioned as frequently as \emph{Neukölln}, which could indicate that the hosts assume that the former does not convey as much meaning as the latter.
However, \emph{Charlottenburg} is often mentioned together with \emph{West Berlin} \textbf{(13)}.
It seems to be important that \emph{Charlottenburg} and particularly the area we have denoted \emph{City West} \textbf{(6)} represents the western center of the formerly divided city:

\begin{quote}
\emph{``Very charming and convenient neighborhood in West-Berlin''} (16763002)\\
\emph{``This is a city center of West Berlin''} (491528)
\end{quote}

\emph{Neukölln} likewise belongs to the former area of \emph{West Berlin}~\cite{SenatskanzleiBerlin2018}, yet, this fact has not been mentioned once.
Generally, \emph{City West} is described as a \emph{nice} and \emph{elegant} area, opposed to the \emph{hip}, \emph{cool}, \emph{trendy}, \emph{vibrant}, and \emph{upcoming} neighborhood \emph{Kreuzkölln} (see Figure~\ref{fig:adjectives}).
The following quote illustrates how hosts construct \emph{City West}:

\begin{quote}
\emph{``Friendly, quiet and trendy area with beautiful old buildings, upscale shops and fine restaurants. People in Charlottenburg are down to earth, educated and wealthy. The typical Berlin Tourism is virtually non-existent. So, perfect for visitors who prefer museums rather than party miles.'' (5921316)}
\end{quote}

This quote indicates that the two analyzed neighborhoods differ greatly in terms of infrastructure and the people they attract.
While \emph{Neukölln} is framed by many Airbnb hosts as \emph{``the place to be''} for a younger crowd, \emph{Charlottenburg} is demarcated from this Berlin image.
As mentioned above, it is mainly depicted as \emph{famous}, \emph{beautiful}, \emph{nice}, and \emph{elegant} (see Figure~\ref{fig:adjectives}).
In particular, these attributes are often used to characterize the main street \emph{Kurfürstendamm} and the historic square \emph{Savignyplatz}.

\begin{table}
\small
\centering
\caption{Places in the city or in the neighborhood, mentioned in the 100 analyzed Airbnb listings.}
\label{tab:places}
\begin{tabular}{ll  rr}
\hline
\multicolumn{1}{c}{\textbf{Category}} & \multicolumn{1}{c}{\textbf{Subcategory}} & \multicolumn{2}{c}{\textbf{Number of listings}} \\
& & \multicolumn{1}{c}{Kreuzkölln} & \multicolumn{1}{c}{City West} \\
\hline
\hline
\textbf{City} & & & \\
& Berlin & 24 & 25  \\
& West Berlin & 1 &\graybg 13 \\
& East Berlin & 1 & 2  \\
\cline{3-4}
& & 24 & 32 \\
\hline
\hline
\textbf{Districts} & & & \\
& Neukölln &\graybg 40 & 1 \\
& Charlottenburg & 1 &\graybg 21 \\
& Kreuzberg &\graybg 15 & 2 \\
\multicolumn{1}{c}{\scriptsize(unofficial)} & Kreuzkölln &\graybg 15 & 0 \\
& Mitte & 5 & 3 \\
\multicolumn{1}{c}{\scriptsize(unofficial)} & City West & 0 & 6 \\
& Other & 4 & 1 \\
\cline{3-4}
& & 46 & 26 \\
\hline
\hline
\textbf{Streets} & & & \\
& Kurfürstendamm & 0 &\graybg 35 \\
& Weserstraße &\graybg 17 & 0 \\
& Wilmersdorfer Straße & 0 & 9 \\
& Kantstraße & 0 & 7 \\
& Other & 9 & 6  \\
\cline{3-4}
& & 21 & 36 \\
\hline
\hline
\textbf{Squares} & & & \\
& Savignyplatz & 0 &\graybg 21  \\
& Alexanderplatz & 7 & 1 \\
& Hermannplatz & 5 & 0 \\
& Other & 5 & 4\\
\cline{3-4}
& & 13 & 24 \\
\hline
\hline
\textbf{Sights} & & & \\
& Berlin Zoo & 1 &\graybg 13 \\
& Opera (German/State) & 0 &\graybg 10 \\
& KaDeWe & 0 &\graybg 8 \\
& Charlottenburg Palace & 0 & 5 \\
& Fair/Messe & 0 & 4 \\
& Brandenburg Gate & 1 & 2 \\
& Other (neighborhood) & 1 & 7 \\
& Other (city) & 2 & 1 \\
\cline{3-4}
& & 3 & 27 \\
\hline
\hline
\textbf{Parks} & & & \\
& Canal/Maybachufer &\graybg 28 & 0 \\
& Görlitzer Park & 6 & 0 \\
& Tiergarten & 0 & 4 \\
& Tempelhofer Feld & 4 & 0 \\
& Lake Lietzensee & 0 & 4 \\
& Other & 1 & 2 \\
\cline{3-4}
& & 32 & 10 \\
\hline
\hline
\end{tabular}
\end{table}

\begin{table}
\small
\centering
\caption{Facilities and people in the neighborhood, mentioned in the 100 analyzed Airbnb listings.}
\label{tab:facilities}
\begin{tabular}{ll  rr}
\hline
\multicolumn{1}{c}{\textbf{Category}} & \multicolumn{1}{c}{\textbf{Subcategory}} & \multicolumn{2}{c}{\textbf{Number of listings}} \\
& & \multicolumn{1}{c}{Kreuzkölln} & \multicolumn{1}{c}{City West} \\
\hline
\hline
\textbf{Transportation} & & & \\
& Public, sharing, etc. & 42 & 37  \\
\hline
\hline
\textbf{Gastronomy} & & & \\
& Restaurants &\graybg 32 &\graybg 25  \\
& Cafés &\graybg 27 & 13  \\
& Bars and Pubs &\graybg 37 & 9 \\
\cline{3-4}
& & 40 & 25 \\
\hline
\hline
\textbf{Shopping} & & & \\
& Non-grocery &\graybg 18 & \graybg 22  \\
& Grocery & 10 & 11 \\
& Weekly markets &\graybg 13 & 5 \\
\cline{3-4}
& & 25 & 23 \\
\hline
\hline
\textbf{Culture} & & & \\
& Art/Galeries &\graybg 11 & 3 \\
& Night clubs &\graybg 12 & 2 \\
& Opera & 0 &\graybg 8  \\
& Cinemas & 1 & 3 \\
& Theaters & 2 & 2 \\
& Other & 3 & 3 \\
\cline{3-4}
& & 21& 16 \\
\hline
\hline
\textbf{People} & & & \\
& Background, age, etc. & 8 & 4  \\
\hline
\hline
\textbf{Parks} & & & \\
& See places (Table~\ref{tab:places}) &\graybg 32 & 9  \\
\hline
\hline
\textbf{Other} & & & \\
& Spa, ATM, etc. & 2 & 8  \\
\hline
\hline
\end{tabular}
\end{table}

\subsection{Streets and Squares}
\label{sec:streets-squares}

Large streets and squares structure the urban fabric~\cite{Levy1999} of a neighborhood.
In our case study, for each neighborhood one important street emerged from the listing descriptions.
\emph{Kurfürstendamm} is the street that is mentioned most frequently \textbf{(35)} and it is primarily framed as Berlin’s \emph{famous} shopping street: \emph{``Berlin’s main shopping street Kurfürstendamm offers you Berlin’s widest choice of shops, from the standard clothes chains to the exclusive designer store''} (989871).
\emph{Weserstraße} is \emph{Kurfürstendamm}’s counterpart in \emph{Kreuzkölln}.
Yet, the street's name is not mentioned in as many listing descriptions as \emph{Kurfürstendamm}, which could indicate that the street is not as popular yet.

Compared to \emph{Kurfürstendamm}, \emph{Weserstraße} is described very differently:
Contrary to the practice of \emph{shopping} encouraged by Airbnb hosts in \emph{City West}, \emph{Weserstraße} mainly provides gastronomic infrastructure, in particular the trilogy of \emph{restaurants}, \emph{cafés}, and \emph{bars/pubs} is mentioned (see Table~\ref{tab:facilities}).
In total, 40 out of 50 analyzed listings in \emph{Kreuzkölln} referred to such facilities, opposed to 25 listings in \emph{City West}.
Hosts did not only mention the prevalent infrastructure as assets of the neighborhood, they likewise inform related practices such as \emph{``enjoy amazing breakfast, lunch and delightful dinners''} (912787), \emph{``have a beer''} (10005680) in the numerous bars and restaurants filled with \emph{``young people''} (7738887), and \emph{``enjoy a typical Berlin summer night''} (10005680).

The most frequently named square to go out for food and drinks in \emph{City West} is \emph{Savignyplatz} \textbf{(21)}.
However, in that neighborhood hosts focused more on \emph{restaurants} than on \emph{cafés} and \emph{bars} (see Table~\ref{tab:facilities}).
\emph{Savignyplatz} was described as an \emph{``upscale part of town''} (16763002) that is \emph{``legendary and historic''} (16763002), offering \emph{``nice restaurants''} (6931794) in an environment with \emph{``flair and ambience''} (6931794).
\emph{Weserstraße}, in contrast, was framed as \emph{``one of the most trendy and lively streets of Neukölln''} (1559030) that has \emph{``some of the coolest bars in Berlin''} (7616493). 
Hence, Airbnb hosts not only contribute to a discursive image construction of a neighborhood, they also advise their guests about what to do and how to behave---they describe adequate place performances.

The location of the offered room or apartment was frequently mentioned in both areas.
Two aspects seem to be particularly important (see Figure~\ref{fig:adjectives}):
First of all, being located in a \emph{central} part of Berlin was mentioned by many hosts in \emph{City West} \textbf{(17)}.
Interestingly, this attribute was rarely used by \emph{Kreuzkölln} hosts \textbf{(3)}, despite the fact that both neighborhoods have approximately the same walking distance to the \emph{Brandenburg Gate} (about 4 km).
The second aspect is being located in a (relatively) \emph{quiet} street of the neighborhood, which was also mentioned more frequently in \emph{City West} compared to \emph{Kreuzkölln} \textbf{(11 vs. 7)}.

\subsection{Sights, Parks, and Markets}
\label{sec:sights-parks-markets}

Sights and other attractions, such as museums or art galleries, are often considered to be facilities that are primarily tourism-related and attract visitors.
In Ashworth and Tunbridge’s \emph{tourist-historic city model}~\cite{AshworthTunbridge1990, AshworthTunbridge2000}, they belong to the decisive elements that define urban tourism space.
This section deals with sights that were mentioned in Airbnb listings.
We evaluate the relevance that traditional sights have in the descriptions and analyze what hosts signify to be worth visiting in case their neighborhood does not provide any major sights. 

Several hosts in \emph{City West} refer to internationally known sights that are located in close proximity to their apartments. 
They frequently name the \emph{Berlin Zoo} \textbf{(13)}, the \emph{opera houses} \textbf{(10)}, the \emph{KaDeWe} \textbf{(8)}, which is one of Europe’s largest department stores, or the \emph{Charlottenburg Palace} \textbf{(5)}.
All of these sights, apart from the \emph{State Opera}, are located in the \emph{City West} neighborhood. 
The \emph{Kaiser Wilhelm memorial church}, which is a landmark building dominating the \emph{Breitscheidplatz} square, located at the beginning of \emph{Kurfürstendamm}, was only mentioned in two listing descriptions.
Other famous Berlin sights such as Reichstag, Television Tower, or Gendarmenmarkt~\cite{visitBerlin2018} were not mentioned in any of the listings. 

Another result of our analysis is that only sights in close proximity to the offered room or apartment seem to be relevant for the hosts to describe. 
While street names, in particular \emph{Kurfürstendamm}, are mentioned in 36 out of 50 analyzed listings in \emph{City West}---\emph{Berlin Zoo}, the most frequently mentioned sight, appears only in 13 listing descriptions (see Table~\ref{tab:facilities}).
Even more striking, is the fact that references to well-known tourist attractions are almost non-existent in \emph{Kreuzkölln}'s Airbnb listings.
As discussed earlier, the neighborhood is de facto lacking major sights.
Nevertheless, \emph{visitBerlin}, the city’s DMO, promotes, for example, the former area of the \emph{Kindl brewery}, which is now transformed into a center for contemporary art~\cite{visitBerlin2018b}.
In addition, \emph{Britz Castle}~\cite{visitBerlin2018c} and the \emph{Hufeisensiedlung}~\cite{visitBerlin2018d}, an UNESCO world heritage site, are promoted.
However, these sites did not appear in any of the \emph{Kreuzkölln} Airbnb listing descriptions we analyzed.

Instead of such traditional tourist sights, hosts framed other facilities as being relevant for visitors.
In particular, the small waterway \emph{Landwehrkanal} at the northern border of \emph{Kreuzkölln} was marked to be important.
This is interesting, because \emph{Landwehrkanal} is not a park offering large green spaces or street furniture. 
It is primarily a canal set in concrete, surrounded by a cobblestone street at the \emph{Neukölln} side and a bicycle path in \emph{Kreuzberg}.
The quality and accessibility of these streets are so bad that the city district office of Friedrichs\-hain\--\-Kreuz\-berg is planning to renovate the area~\cite{BezirksamtFriedrichshainKreuzberg2018}.
For local Airbnb hosts, however, this facility is an attraction, it is framed as a \emph{beautiful} and \emph{famous} place:  

\begin{quote}
\emph{``The place is next to the beautiful canal in Neukölln''} (7738887)\\
\emph{``Right next door is the famous `canal' that separates Kreuzberg from Neukölln''} (2246227)
\end{quote}

Furthermore, the place names \emph{Landwehrkanal} and \emph{Maybachufer} (the banks of the canal) appeared in more than half of the analyzed listings in \emph{Kreuzkölln}.
This high amount of namings indicates that Airbnb hosts regard this facility as significant for visitors.
However, \emph{Landwehrkanal} is hardly a place that can be experienced through the typical tourist practice of \emph{gazing}.
According to Urry’s seminal work \emph{``The Tourist Gaze''}~\cite{Urry1990}, the practice of gazing upon landscapes, places, objects, or people is a defining element of tourism.
It is closely related to the practice of sightseeing that is common in classic city tourism.
Instead of gazing upon \emph{Landwehrkanal}, hosts encouraged other practices.
They invited guests to \emph{``take a bike ride along the tree-lined Landwehr Canal''} (4243519), to \emph{``stroll''} along (6089778), to \emph{``hang out''} (879602), or to \emph{``jog or take walks''} (16462941).
One host even related being at the canal to a common local practice: \emph{``(...) around the corner (...) is the canal where you can have a beer and enjoy a typical Berlin summer night''} (10005680).
Highlighting that relaxing at \emph{Landwehrkanal} is a typical local practice enables visitors to behave in a similar way.
They learn how to perform in order to immerse themselves in the local everyday life and to deliberately blur the boundaries between visitor and resident.

Referring to places such as \emph{Maybachufer}, \emph{Tempelhofer Feld} \textbf{(4)}, or \emph{Görlitzer Park} \textbf{(6)} in the Airbnb listings guides visitors to find places that are mainly frequented by local residents for their leisure activities.
Generally, they are not worth visiting due to their physical appearance---\emph{Tempelhofer Feld} might be an exception due to its size.
Instead, it is the way hosts describe these areas that makes them attractive for visitors who want to experience Berlin like a local:

\begin{quote}
\emph{``One of my favorite places in Berlin [is] the Tempelhof airfield, [...] a lovely park''} (1590153)\\
\emph{``Tempelhofer Feld, the former aircraft field that has been transformed into a marvelous park''} (5963351) 
\end{quote}

These findings, however, cannot be transferred to all parks mentioned in the analyzed listings.
In \emph{City West} listing descriptions, hosts referred to the large and famous \emph{Tiergarten} \textbf{(4)} and \emph{Lake Lietzensee} \textbf{(4)}.
However, only one host encouraged the practice of \emph{``relaxing''} (2141544).
In all other cases, parks are mentioned together with other sights, particularly in the case of \emph{Tiergarten}.
Despite its size of 210 hectares~\cite{visitBerlin2018e}, and its location in the center of Berlin, adjoining \emph{Kurfürstendamm}, not much attention was given to it in the Airbnb listings. 

In line with \emph{parks} being described as everyday facilities worth visiting, local \emph{markets} and regular \emph{grocery} shopping gains importance.
\emph{Gastronomic} offers and \emph{shopping} facilities are of relevance in both neighborhoods, although their framings vary.
While attention in \emph{City West} lies on expensive and exclusive \emph{designer stores} (989871, 846489), cheap and \emph{second-hand bargains} (8376421) are promoted in \emph{Kreuzkölln}.
In particular, the large department store \emph{KaDeWe} \textbf{(8)} is mentioned as a shopping paradise and as an attraction to visit in \emph{City West}.
The corresponding facilities in \emph{Kreuzkölln} are the immigrant food market and the bi-weekly flea market \textbf{(13)} at \emph{Maybachufer}.
This food market, which originally satisfied the needs of the large immigrant population in \emph{Kreuzberg} and \emph{Neukölln}~\cite{AmtfurStatistikBerlinBrandenburg2017e}, is now being marked as a multicultural place for grocery shopping:

\begin{quote}
\emph{``very rich and colorful turkish market at Maybachufer, where you can buy fresh fish, meet, turkish nuts and dried fruits etc.''} (1590153)\\   
\emph{``You will find a huge open air grocery market twice a week (extra good bio products) and an amazing flea market every second Sunday''} (7738887)
\end{quote}

In these listing descriptions, the history of the neighborhood \emph{Kreuzkölln} is prevalent.
Airbnb hosts are steering tourists’ attention towards the market, which is rooted in the neighborhood.
They highlight its long tradition and the fact that it is still frequented by both long established as well as short-term inhabitants.   
This mix of people visiting it, and the offers that are still primarily dedicated to the local residents, mark it as an authentic place---a place that is not artificially designed to meet only visitors' needs.

\begin{table}
\small
\centering
\caption{Places mentioned by \emph{visitBerlin}, the city's DMO, in their borough descriptions of \emph{Neukölln} and \emph{Charlottenburg-Wilmersdorf}, the boroughs in which the two neighborhoods \emph{Kreuzkölln} and \emph{City West} are located; in brackets, we provide number and percentage of Airbnb listings in the corresponding neighborhood (\emph{Kreuzkölln}: $n=750$, \emph{City West}: $n=210$) mentioning the place; asterisk indicates places that were only used to describe the location of other places.}
\label{tab:places-dmo}
\begin{tabular}{l rr}
\hline
 & \multicolumn{2}{c}{\textbf{Places from DMO borough descriptions and matches in Airbnb listings}} \\
\multicolumn{1}{c}{\textbf{Paragraph}} & \multicolumn{1}{c}{Neukölln} & \multicolumn{1}{c}{Charlottenburg-Wilmersdorf} \\
\hline
\hline
\multirow{3}{*}{\textbf{Header}} &\graybg Neukölln (district) (468 | 62.4\%) &\graybg Charlottenburg (district) (90 | 42.9\%)  \\
& &\graybg Wilmersdorf (district) (43 | 20.5\%) \\
& &\graybg City West (neighborhood) (16 | 7.6\%)\\
\hline
\hline
\multirow{5}{*}{\textbf{Favorite places}} & Richardplatz (square) &\graybg Bikini Berlin (concept mall) (19 | 9.0\%)  \\
& Schloss Britz/Gutspark (palace/park) &\graybg A-Trane (jazz club) (7 | 3.3\%)\\
& Capt‘n Crop (hat making studio) &\graybg Teufelsberg (hill) (1 | 0.5\%)\\
& Kindl (art space) & Bröhan Museum (museum)\\
& & Rüdesheimer Platz (square)\\
\hline
\hline
\multirow{4}{*}{\textbf{Need to know}} &\lightgraybg Kreuzberg (227, 30.3\%)* &\graybg Kurfürstendamm (street)  (119 | 56.7\%)  \\
&\graybg Schillerkiez (neighborhood) (5 | 0.7\%) &\graybg Grunewald (district/park) (3 | 1.4\%)\\
& Gropiusstadt (neighborhood) & \\
& Lavanderia Vecchia (restaurant) & \\
& Eins 44 (restaurant) & \\
\hline
\hline
\multirow{4}{*}{\textbf{Other}} & \graybg Weserstraße (street) (189 | 25\%) &\graybg  Kurfürstendamm (street)  (119 | 56.7\%) \\
& \lightgraybg Landwehr Canal (canal) (92 | 12\%)* &\graybg West Berlin (part of city) (28 | 13.3\%) \\
& \lightgraybg Sonnenallee (street) (62 | 8.3\%)* &\graybg Charlottenburg Pal. (palace/park) (20 | 9.5\%) \\
& \lightgraybg Hermannstraße (street) (7 | 0.9\%)* &\graybg Bikini Haus (mall) (19 | 9.0\%) \\
&\graybg Tier (bar) (3 | 0.4\%) &\graybg City West (neighborhood) (16 | 7.6\%) \\
&\graybg Ä (pub) (1 | 0.1\%) &\graybg Olympic Stadium (stadium/tower) (5 | 2.4\%)\\
&\graybg Körner Park (park) (1 | 0.1\%) &\graybg Uhlandstraße (street) (7 | 3.3\%) \\
& Heimathafen (arts/theater) &\graybg East Berlin (part of city) (3 | 1.4\%) \\
& Neuköllner Oper (opera) &\graybg Grunewald (district/park) (3 | 1.4\%) \\
& Vin Aqua Vin (bar) &\graybg Waldorf Astoria (hotel) (2 | 1.0\%) \\
& Schloss Britz (palace) &\graybg Funkturm (radio tower) (2 | 1.0\%) \\
& Britzer Garten (park) & \graybg Wannsee (lake) (2 | 1.0\%) \\
& & \graybg Teufelsberg (hill) (1 | 0.5\%) \\
& & Schloßstraße (street) \\
& & Rüdesheimer Straße/Platz (street/square) \\
& & Fasanenplatz (square) \\
& & Ludwigkirchplatz  (square) \\
& & Haus der Berliner Festspiele (theater) \\
& & Bar jeder Vernunft (theater) \\
& & Havel (river) \\
& & Waldbühne (amphitheater) \\
& & Grunewaldturm (tower) \\
\hline
\hline
\end{tabular}
\end{table}

\subsection{Comparison with DMO Website}
\label{sec:comparison-dmo}


After extracting the places mentioned in \emph{visitBerlin}'s description of the two boroughs in which the neighborhoods \emph{Kreuzkölln} and \emph{City West} are located (see Section~\ref{sec:data-collection}), we first compared those places with the places frequently mentioned by Airbnb hosts (gray background in Table~\ref{tab:places}).
Then, we conducted a follow-up quantitative analysis to search for the places mentioned by the DMO in all 960 Airbnb listing descriptions we retrieved (see Section~\ref{sec:quantitative-analysis} for a description of our search approach).
Table~\ref{tab:places-dmo} shows the results of this analysis.

Regarding the district names, it was more important for hosts in \emph{Kreuzkölln} to refer to the district name \emph{Neukölln} as it was for hosts in \emph{City West} to refer to \emph{Charlottenburg} or \emph{Wilmersdorf} (those two districts were merged in 2001 and \emph{City West} is part of former \emph{Charlottenburg}).
We also observed this trend in our qualitative analysis (see Section~\ref{sec:places-facilities}).

Regarding the DMO's favorite places, it is noteworthy that none of the four favorite places in \emph{Neukölln} were mentioned by the Airbnb hosts in \emph{Kreuzkölln}, while hosts in \emph{City West} named three out of five places in \emph{Charlottenburg-Wilmersdorf}.
The same trend can also be observed for the other sections of the descriptions, where the overlap between the DMO and the hosts in \emph{City West} was relatively large.
Interestingly, while \emph{Neukölln}'s neighboring district \emph{Kreuzberg} was named in 227 \emph{Kreuzkölln} listings (30.3\%), \emph{visitBerlin} only used the name \emph{Kreuzberg} to describe the location of the \emph{Kreuzkölln} neighborhood, without mentioning its name.
The DMO just referred to the neighborhood as \emph{``a vibrant multicultural area around the border to Kreuzberg''}~\cite{visitBerlin2018h}.
While Airbnb hosts relate their \emph{upcoming} neighborhood \emph{Neukölln} to the already better known \emph{Kreuzberg}, the DMO tries to market the two boroughs independently from each other.

Hosts and \emph{visitBerlin} agree that \emph{Kudamm} and \emph{Weserstraße} are the main streets in the two areas.
The DMO also motivates similar practices like the Airbnb hosts: \emph{shopping} for \emph{Kudamm} and \emph{having a drink at a bar} for \emph{Weserstraße}.
The latter practice was also motivated by the hosts in \emph{City West} for the area around \emph{Savignyplatz}. 
However, the DMO did not even mention this square or the practice of going out for a drink in their description of \emph{Charlottenburg-Wilmersdorf}.

Our qualitative analysis revealed that traditional tourist sights such as \emph{Berlin Zoo} or \emph{KaDeWe} were primarily named by the hosts in \emph{City West}---\emph{visitBerlin} did not mention those sights in their borough description of \emph{Charlottenburg-Wilmersdorf}.
The only obvious overlap between Airbnb hosts and the DMO is \emph{Charlottenburg Palace}. 

The qualitative analysis also suggested that hosts in the new urban tourism hotspot \emph{Kreuzkölln} reframe everyday places such as the \emph{Landwehrkanal}/\emph{Maybachufer}, including its weekly \emph{market}, as being significant for tourists.
The DMO did not mention the \emph{market} or the \emph{Maybachufer}, and only referred to \emph{Landwehrkanal} to describe the location of a different facility.

Generally, the overlap between places mentioned by \emph{visitBerlin} and places mentioned by the Airbnb hosts is much larger in \emph{City West} compared to \emph{Kreuzkölln}, even though we retrieved a considerably smaller number of listings in \emph{City West} (210 vs. 750). 
While hosts in \emph{Kreuzkölln} focused on places in the neighborhood and nearby, which the DMO did not consider to be noteworthy, hosts in \emph{City West} mentioned places all over the district of \emph{Charlottenburg-Wilmersdorf}, which were also named by the DMO.

\section{Discussion}
\label{sec:discussion}


Questions regarding the conceptual nature of space and place have been addressed in both CSCW and human geography research.
For CSCW scholars, the main motivation to use spatial metaphors has long been to transfer spatial structures from the physical world into the digital realm~\cite{HarrisonDourish1996}.
Examples include the organization of the virtual workspace~\cite{BrewerDourish2008} or the creation of new collaborative virtual environments~\cite{HarrisonDourish1996}.
More recent considerations, however, established that space cannot solely be regarded as a \emph{``natural fact---a collection of properties that define the essential reality of settings of action''}~\cite{Dourish2006}; space, like place, is a social product~\cite{Dourish2006}. 

The proliferation of peer-to-peer (information) sharing platforms have influenced the way people encounter (urban) space.
As Brewer and Dourish point out, mobile communication platforms not just produce another level of \emph{``virtual space''} on top of the physical space---they \emph{``allow people to encounter and appropriate existing spaces in different ways''}~\cite{BrewerDourish2008}.
Technology thus becomes a part of how people encounter (urban) space and it \emph{``is shaped through technologically mediated mobility''}~\cite{BrewerDourish2008}.     
Our research draws on these theoretical considerations and expands their implications into the field of leisure and tourism, a research area that already received attention in CSCW contributions~\cite{BrownChalmers2003, Dourish2006}.

In this paper, we investigated how online platforms such as Airbnb engage in the co-production of (new) urban tourism space. 
We considered the \emph{``spatial turn''}~\cite{Soja1989} in social sciences as the starting point for our reflections on the nature of space as being socially constructed.
We then used the \emph{tourist-historic city model}~\cite{AshworthTunbridge1990} to illustrate that traditional framings of urban tourism space as clusters of historical sights, leisure facilities, and gastronomic infrastructure do not suffice to explain the emergence of new urban tourism areas in residential neighborhoods. 
The proliferation of mobile technologies and changed tourist behavior led to rising visitor numbers in \emph{off the beaten track} localities~\cite{MaitlandNewman2009}.
Urban tourism scholars discuss this phenomenon as \emph{new urban tourism}~\cite{FullerMichel2014}.
To explain how residential neighborhoods lacking major sights can gain significance for visitors, we utilized the concepts of \emph{representations}~\cite{Iwashita2003, PritchardMorgan2000, Saarinen2004} and \emph{performances}~\cite{BrenholdtHaldrupOthers2004, Edensor1998, Edensor2000, Edensor2001, Larsen2008} as two theoretical lenses.
We argued that Airbnb hosts publish strategically produced representations of their neighborhoods online and thus alter the spatial discourse~\cite{Davis2005, Saarinen2004}. 
They endow residential neighborhoods with new meanings, encourage place-specific practices, and consequently co-construct places of significance for visitors.
By the means of such collaboratively produced spatial representations, residential neighborhoods transform into (new) urban tourism destinations. 

In our case study, we considered one example of such digital representations.
We qualitatively analyzed  how the two Berlin neighborhoods \emph{Kreuzkölln} and \emph{City West} are digitally constructed by Airbnb hosts in their listing descriptions.
Moreover, we quantitatively investigated to what extend mentioned places differ between the listing descriptions and the digital representation of the corresponding boroughs provided by Berlin’s DMO.
We found that the types of places and sights described greatly differ between the two neighborhoods \emph{Kreuzkölln} and \emph{City West}.
While \emph{Kreuzkölln} hosts mainly reframed everyday places to be worth visiting, \emph{City West} hosts primarily focused on well-known sights that were likewise promoted by the DMO.  

In the neighborhood of \emph{Kreuzkölln}, the places and facilities described to be worth visiting originally served the needs of local residents.
A \emph{canal} bordering \emph{Neukölln} (\emph{Landwehrkanal}) and a weekly \emph{food market} were marked as the areas’ highlights.
These facilities, however, do rarely attract visitors due to their physical appearance.
Instead, Airbnb hosts convey specific meanings to these places.
In their neighborhood descriptions, hosts signify \emph{Landwehrkanal} as a beautiful place to spend a typical Berlin summer night and describe the food market as an exotic place.
Hosts encourage their guests to buy typical food in order to experience the neighborhoods' multicultural atmosphere.
As a result of these neighborhood descriptions, certain space images are collaboratively constructed and (re-)produced by Airbnb hosts.
In the case of \emph{Kreuzkölln}, a residential area and its prevalent local infrastructure is reinterpreted as a touristic place.  

The second neighborhood \emph{City West}, in contrast, provides facilities that are generally understood as sights or attractions.
In their listing descriptions, hosts focused primarily on sights in close proximity to their room or apartment.
Iconic architecture that is located further away, such as \emph{Brandenburg Gate} or \emph{Reichstag}, was rarely mentioned.
Airbnb hosts in \emph{City West} steered visitors' attention mainly to \emph{Kurfürstendamm}, a large street that they signified as a shopping paradise, and to \emph{Savignyplatz}, which they described as a historic square full of restaurants.
Hosts primarily encouraged the practices of \emph{shopping} and \emph{eating out}, which they directly related to the prevalent infrastructure.
They do not need to reinterpret the area, as in the case of \emph{Kreuzkölln}, to mark it as attractive for visitors.
Instead they derive its attractiveness from the neighborhood’s past.
They refer to the famous \emph{Kurfürstendamm} and to \emph{KaDeWe}, one of the largest and oldest department stores in Europe, founded in 1907.
Hence, \emph{City West} hosts reproduce long-established images and practices of their neighborhood in their listing descriptions.

In both cases, the analyzed neighborhoods are signified by hosts as places worth visiting.
While \emph{Kreuzkölln} hosts reinterpret everyday facilities, \emph{City West} hosts reproduce existing space images.
Hence, hosts in both neighborhoods contribute to the perception of their neighborhoods through reinterpreting or reproducing space images---they construct them as tourist places.

The quantitative analysis revealed that the overlap between the DMO's borough descriptions and hosts' neighborhood descriptions is larger in the traditional urban tourism hotspot \emph{City West} compared to the new urban tourism hotspot \emph{Kreuzkölln}.
In both districts, the DMO mainly focused on classic, material sights.
Many of the places located in \emph{Charlottenburg-Wilmersdorf} were also mentioned by Airbnb hosts in \emph{City West}.
In contrast, none of the DMO's favorite places in the borough \emph{Neukölln} were mentioned by \emph{Kreuzkölln} hosts.
Those hosts instead focused on local infrastructure and everyday places such as the \emph{Landwehrkanal}/\emph{Maybachufer} and the \emph{markets}, which the DMO did not consider to be noteworthy.
Interestingly, \emph{Landwehrkanal} appears in \emph{visitBerlin}'s new tourism concept \emph{``Berlin-Tourismus 2018+''}~\cite{DWIFConsulting2017} as a place that is gaining increasing visitor attention.
The authors estimate 27\% of the \emph{Landwehrkanal} visitors to be \emph{``new urban tourists''}.
The DMO's rising awareness regarding \emph{Landwehrkanal} as a tourist attraction indicates that Airbnb hosts' place framings already play a crucial role in steering visitor attention and mobility.

In summary, our findings show how space images are collaboratively constructed in Airbnb listings and how they differ between the analyzed neighborhoods.
We found that Airbnb hosts in traditional tourist hotspots rather reproduce existing place images, which have previously been co-produced by the city's DMO.
In contrast, hosts in new urban tourism hotspots tend to reinterpret everyday places and endow them with new meanings, but such places are often not regarded as being significant for visitors by the DMO.
The new urban tourism hotspot \emph{Kreuzkölln}, for example, seems to provide a variety of attractions according to Airbnb hosts, but is not promoted as a touristic place by Berlin's DMO.
Since the DMO seems to focus on material sights in general, they are unlikely to promote residential areas that are lacking such infrastructure.
On peer-to-peer platforms like Airbnb, in contrast, everyday places gain importance and are marketed.
We thus hypothesize that digital information sharing platforms are more important in the production of new urban tourism areas than the city's DMO. 

Our finding that, through digital representations of space in Airbnb listing descriptions, mundane places can gain significance for visitors and thus transform into tourist places, is an important contribution, because previous research mainly focused on spatial representations and destination images produced by the city’s DMO or other governmental representatives~\cite{ChenChen2016, PritchardMorgan2000}.
We, in contrast, have illustrated how space images can likewise be produced and reframed collaboratively on online platforms such as Airbnb.
Against this background, we argue that the power to endow space with new meanings is nowadays more evenly distributed among actors.
The DMO or the tourism industry in general are no longer the only ones pre-interpreting tourism space.
By the means of digital technologies and online platforms, local people can likewise contribute to the framing of places.

A last aspect we want to discuss here is how digital technologies encourage particular appropriations of space.
In our case study, we analyzed performances motivated in Airbnb listings.
For example, we found that Airbnb hosts in both neighborhoods encourage the practice of \emph{going out} in their listing descriptions.
Moreover, while hosts in \emph{City West} rather focused on the practice of \emph{shopping}, likely motivated by the presence of \emph{KaDeWe} and \emph{Kurfürstendamm} in close proximity, Airbnb hosts in \emph{Kreuzkölln} encouraged their guests to \emph{relax} at the shore of \emph{Maybachufer} or to \emph{enjoy the nightlife} in one of the various bars and restaurants nearby.
The practice of \emph{sightseeing}, which is traditionally regarded as a typical tourist activity, plays a subordinate role.   
These findings illustrate that Airbnb hosts not only reproduce or reinterpret spatial representations in their listings, but also influence the way space is enacted through the practices they encourage.
As a direction for future work, we suggest to add an ethnographic perspective to our research design in order to investigate how exactly places are enacted by Airbnb guests.

\section{Conclusion and Future Work}


In this paper, we illustrated how urban tourism space is (re-)produced digitally and collaboratively on online platforms.
In particular, we investigated how Airbnb hosts construct their neighborhoods as touristic places in their listing descriptions.
We followed a constructionist notion of space, building on existing research in human geography~\cite{Soja2009} and CSCW~\cite{HarrisonDourish1996, Dourish2006}.
We understand Airbnb listing descriptions as \emph{representations} of space~\cite{Iwashita2003, PritchardMorgan2000, Saarinen2004}, produced by Airbnb hosts and read by potential guests, which have the power to influence the discourse about an area and the way places are appropriated.   

For our empirical study, we collected Airbnb listing data from the two Berlin neighborhoods \emph{Kreuzkölln} and \emph{City West} and qualitatively analyzed a random sample of 100 listing descriptions.
We found that, in the description of their neighborhood, hosts primarily focused on facilities in close proximity to their apartment.
Well-known sights that are further away were of little importance.

In the neighborhood \emph{Kreuzkölln}, which is basically lacking any major sights, hosts reframed local everyday places as being of significance for visitors.
The shores of \emph{Maybachufer}, a canal bordering \emph{Neukölln}, and a weekly food \emph{market} were framed as the areas' highlights.
These facilities are no sights in a traditional sense.
Instead, Airbnb hosts reinterpret such mundane places and convey new meaning to them.
Our qualitative analysis of the listings in the \emph{City West} neighborhood revealed that hosts mainly focused on traditional sights, such as \emph{Berlin Zoo} and \emph{KaDeWe}, and motivated related practices.  

Our quantitative analysis has shown that the space construction between Airbnb hosts and the DMO is more similar in traditional tourism hotspots (like \emph{City West}) compared to new urban tourism hotspots (like \emph{Kreuzkölln}).
We conclude that online platforms such as Airbnb play a crucial role in (re-)directing visitors' attention to less `touristified' neighborhoods.
Moreover, we illustrated how encouraged place-specific \emph{performances} help visitors to appropriate such neighborhoods.

Our research approach opens up various directions for future work.
Particular projects include analyzing the construction of tourism space in Airbnb listings located in other neighborhoods or cities, but it also seems promising to compare how tourism space is constructed on other platforms such as Instagram or TripAdvisor.
Another direction would be to scale the approach we followed in our quantitative analysis, comparing the DMO's descriptions of all twelve boroughs to all Airbnb listing descriptions in Berlin.
That way, one could test the hypothesis if the DMO's descriptions are more likely to match the Airbnb listing descriptions in traditional urban tourism hotspots like \emph{City West}.
In our case study, the Airbnb listing density was much higher in the new urban tourism hotspot \emph{Kreuzkölln}.
One could use this information, together with other data retrieved from Airbnb listings, to classify neighborhoods, similar to Venerandi et al.'s approach~\cite{VenerandiQuattroneOthers2015}.
An open question that could be investigated using surveys and interviews is if hosts are aware of their role as producers of space, in particular in new urban tourism hotspots. 

On a more general level, understanding tourism space as being socially constructed through diverse forms of \emph{representations} and \emph{performances} enables researchers to further investigate the discourse about a place.
This aspect is becoming increasingly important in light of the proliferation of sharing platforms for different kinds of information such as images and films.
Traditionally, the destination’s management and marketing organization was the dominant actor steering tourists' attention and action spaces in the city.
Sharing platforms such as Airbnb, Instagram, and TripAdvisor are now overtaking this role.
Thus, people producing digital content on these platforms have nowadays a large influence on how certain places are perceived.
Considering this development, it appears to be very promising to foster collaboration between social and information sciences in order to understand how digital media impacts our perception of reality.


\bibliographystyle{ACM-Reference-Format}
\bibliography{literature} 

\end{document}